\documentclass[iopart,twocolumn,superscriptaddress,bibliography]{revtex4-2}

\usepackage{graphicx}
\usepackage{array}
\usepackage{dcolumn}
\usepackage{bm}
\usepackage{float}
\usepackage{enumitem}
\usepackage{longtable}
\usepackage{tablefootnote}
\usepackage{commath}
\usepackage[colorinlistoftodos, textwidth=4cm, shadow]{todonotes}
\usepackage{natbib}

\usepackage{longtable}
\usepackage[utf8]{inputenc}

\begin{document}

\title{Crystal Structure and Magnetism of Actinide Oxides: A Review}

\author{Binod K. Rai}
\email[]{binod.rai@srnl.doe.gov}
\affiliation{Savannah River National Laboratory, Aiken, SC, 29808  USA}

\author{Alex Bretaña} 
\affiliation{Savannah River National Laboratory, Aiken, SC, 29808  USA}

\author{Gregory Morrison}
\affiliation{Department of Chemistry and Biochemistry, University of South Carolina, 631 Sumter Street, Columbia, SC, USA}

\author{Ryan Greer}
\affiliation{Savannah River National Laboratory, Aiken, SC, 29808 USA}

\author{Krzysztof Gofryk}
\affiliation{Idaho National Laboratory, Idaho Falls, Idaho 83402 USA}
\affiliation{Glenn T. Seaborg Institute, Idaho National Laboratory, Idaho Falls, Idaho 83402 USA}

\author{Hans-Conrad zur Loye}
\affiliation{Department of Chemistry and Biochemistry, University of South Carolina, 631 Sumter Street, Columbia, SC, USA}

\date{\today}

\begin{abstract}
In actinide systems, the 5f electrons experience a uniquely delicate balance of effects and interactions having similar energy scales, which are often difficult to properly disentangle. This interplay of factors such as the dual nature of 5f-states, strong electronic correlations, and strong spin-orbit coupling results in electronically unusual and intriguing behavior such as multi-k antiferromagnetic ordering, multipolar ordering, mott-physics, mixed valence configurations, and more. Despite the inherent allure of their exotic properties, the exploratory science of even the more basic, binary systems like the actinide oxides has been limited due to their toxicity, radioactivity, and reactivity. In this article, we provide an overview of the available synthesis techniques for selected binary actinide oxides, including the actinide dioxides, sesquioxides, and a selection of higher oxides. For these oxides, we also review and evaluate the current state of knowledge of their crystal structures and magnetic properties. In many aspects, substantial knowledge gaps exist in the current body of research on actinide oxides related to understanding their electronic ground states. Bridging these gaps is vital for improving not only a fundamental understanding of these systems but also of future nuclear technologies. To this end, we note the experimental techniques and necessary future investigations which may aid in better elucidating the nature of these fascinating systems.

\end{abstract}

\maketitle

ABBREVIATIONS

\section{Glossary of acronyms}\label{sec1}




$fcc$: Face‐Centered Cubic

$bcc$: Body‐Centered Cubic

AFM: Antiferromagnetic

FM: Ferromagnetic


$T_N$: N\'{e}el temperature




$T_0$: Transition temperature

mK: Milli‐Kelvin

$m$g: Milli‐gram

$\mu$g: Micro‐gram

JT: Jahn–Teller

DFT: Density Functional Theory



INS: Inelastic Neutron Scattering

NMR: Nuclear Magnetic Resonance

 
XPS: X-ray Photoelectron Spectroscopy

$\mu$SR: Muon Spin Relaxation



\section{Introduction}\label{sec2}

The consequences of 5$f$ electrons in the actinide elements are as interesting as they are complex, with actinide systems displaying dual nature of 5$f$ states (itinerant-localized), competing interactions, and strong spin-orbit coupling\cite{santini2009RMP,moore2009RMP,marsh2017DT,white2019CEJ}. Actinide materials can have both local and itinerant moment character due to their partially filled 5$f$ shells\cite{zwicknagl2003JPCM}, effects which can exhibit profound competition in actinide compounds. This interplay results in unusual behavior from a magnetism perspective, such as hidden order\cite{tokunaga2006PRL,kung2015Science,carretta2010PRL,caciuffo2011PRB,allen1968PRa,allen1968PRb,faber1975PRL,faber1976PRB,santini2009RMP}, complex magnetic order\cite{tokunaga2014PRB, tokunaga2010JPSJ,prots2022PRB}, topological phenomena\cite{ivanov2019PRX,zhang2012Science}, mixed valence configurations\cite{morss2006Springer,huang2020PRB,desgranges2016IC,schelter2008ACIE}, changes in covalency\cite{pace2021CAEJ,gaiser2021Naturecomm,cary2015Naturecomm}, and Mott transitions\cite{zhang2012Science,moore2009RMP,conradson2013PRB}. Such dual moment nature and large spin-orbit coupling observed in actinide materials contribute to their diverse physical and chemical properties, and therefore, a better understanding of these properties has the potential to advance fundamental actinide science as well as the future of energy and information technology. However, the exploratory science of actinide materials has been extremely limited due to the toxicity, radioactivity, and reactivity of the actinide elements. Consequently, not all established approaches to sample synthesis can be employed, particularly when it comes to single crystals. In addition to challenges related to sample quality, the complexity of actinide oxides is further exacerbated by the absence of an accurate description of their electronic structure. This inaccuracy stems from intricate interactions, including strong electron correlations, crystal field interactions resulting from the electric field generated by surrounding ligands or ions, and coupled spin-orbit interactions that arise from relativistic effects, which intertwine an atom's electron spin and orbital motion. These interactions profoundly impact the electronic structure, magnetic and spectroscopic properties of actinides, influencing their chemical reactivity and bonding patterns. A detailed description of these interactions is well discussed in references\cite{hotta2006ROPP,santini2009RMP,moore2009RMP} and recent advancements in theoretical calculations and modeling which consider these interactions in actinide oxides are discussed here\cite{zhou2022PRB, shiina2022JPSJ,Pang2013PRL,wen2013ChemRev,pegg2018PCCP,huang2020PRB}.

Historically, the bulk of actinide oxide research has focused on the properties of various uranium and plutonium compounds as they impact their use in the nuclear fuel industry. While the fundamental understanding of compounds such as UO$_2$ has improved, a fundamental grasp of the physical properties of other actinide oxides remains elusive. Several synthetic, structural and characterization challenges have impeded a better understanding of the actinide oxides: (i) a dearth of high-quality polycrystalline samples or single crystals, (ii) conflicting crystal structures of the higher actinide oxides, such as Pa$_2$O$_5$, (iii) a severe lack of experimental characterization of many actinide oxides, such as Cm$_2$O$_3$, and (iv) inconsistent and sometimes contradictory experimental magnetic properties, as observed in NpO$_2$ and AmO$_2$.

Synthesizing high-quality samples of the higher actinide oxides, especially as single crystals, has been challenging due to the coexistence of different oxides with different oxidation states, as well as the inherent difficulties in handling the toxic, radioactive, and reactive actinide elements.  This lack of high-quality samples has limited the determination of exact crystal structures of many higher actinide oxides.  While the actinide dioxides crystallize in the simple face-centered cubic ($fcc$) fluorite structure, the sesquioxides and the higher oxides have more complicated crystal structures and can even crystallize in several polymorphic structures. Inadequate crystal quality, along with factors such as the inability of facilities to handle analytically amenable quantities of material and the lack of access to necessary experimental and analytical techniques have drastically impacted investigations into the electronic and magnetic properties, and has made determination of magnetic ground states an enduring problem in actinide materials. This is exemplified in compounds such as AmO$_2$, CmO$_2$, U$_2$O$_5$, and U$_3$O$_7$, whose ground states are still unresolved due to a lack of high-quality samples. M\"{o}ssbauer spectroscopy\cite{kalvius1969PLB} and neutron diffraction measurements have failed to detect any magnetic order below the transition temperature $T_0$ = 8.5 K in AmO$_2$\cite{boeuf1979JDPL}. However, recent Nuclear Magnetic Resonance (NMR) measurements on AmO$_2$ performed by $Tokunaga$ $et$ $al.$\cite{tokunaga2014PRB,tokunaga2010JPSJ} confirmed the phase transition seen at $T_0$. Magnetic characterizations of higher actinide oxides like U$_2$O$_5$, U$_3$O$_7$, and Pa$_2$O$_5$ have yet to be explored. To overview such recent discoveries and uncover the potential of unexplored actinide oxides, there is a need for a review article that summarizes the behavior of actinide oxides.

In this article, we summarize the main challenges in actinide oxide synthesis together with their magnetic properties and the scientific potential of unexplored actinide oxides. More comprehensive reviews on actinide metals, intermetallic actinide materials, and thin film actinide oxide materials have been covered elsewhere\cite{lashley2005PRB,stewart2001RMP,pfleiderer2009RMP,mydosh2011RMP,brian2010JLTP,vallejo2022ROPP,hurley2021CR}. It is not our intention to cover any specialized techniques that have been discussed in detail in other articles, such as NMR\cite{walstedt2014CRP,martel2014IC}, Synchrotron Radiation\cite{shi2014AM}, Density Functional Theory (DFT)\cite{roy2019BOOK}, or X-ray Photoelectron Spectroscopy (XPS)\cite{teterin2004RCR}. For most bulk binary actinide oxides, we cover their crystal growth methods in Section \ref{sec3}, the current state of our understanding of their magnetic properties in Section \ref{sec4}, and summary and recommendations for future work in Section \ref{sec5}.

\section{Crystal Growth}\label{sec3}

Actinides are generally difficult to handle due to their scarcity, reactivity, radioactivity, and toxicity. Light elements, such as Th and U, can be handled in kilogram quantities, but heavier elements are typically limited to $m$g or even $\mu$g sample sizes due to their high activity. Given these limitations, work on actinide materials synthesis is limited, and only very few specialized laboratories across the globe are able to undertake such experiments.  Several reviews of the synthesis of binary actinide oxides exist\cite{haire1994BOOK,morss1991BOOK,katz2007BOOK}.  We have attempted to give an overview of all the synthesis techniques including those developed since the last review over 15 years ago in Table \ref{table1}, along with the corresponding references. A brief description of these methods with selected references is provided below. Structural information for each compound is provided in Table~\ref{table2} and \ref{table3}. The lattice parameters reported are at room temperature unless otherwise stated in the text. 

Polycrystalline samples of actinide oxides are most commonly synthesized by calcination of various precursors including the oxalates\cite{qi2022IJACT,nave1983PRB}  nitrates\cite{vita1973ACA}  and hydroxides\cite{richter1987JNM}. Controlling the atmosphere during calcination, or post-synthetic oxidation/reduction, allows for control over the specific actinide oxide produced.  For example, curium oxides were synthesized by the calcination of curium oxalate.  Cm(III) was precipitated from a 0.1 M HCl solution with oxalate ions.  The resulting curium oxalate was calcined in oxygen at 1000 $^{\circ}$C, resulting in an intermediate oxygen stoichiometry.  This product was then heated at 1100 $^{\circ}$C in vacuum or carbon monoxide to produce Cm$_2$O$_3$, or at lower temperatures in $4-5$ atm of oxygen to produce CmO$_{2}$\cite{nave1983PRB}.  For smaller scale, $\mu$g reactions, calcination of ion-exchange resins is also possible\cite{Baybarz1972aUSatom}. Synthesis of polycrystalline samples of actinide oxides have also been reported through sol-gel\cite{wymer1965preparation} and aqueous routes\cite{kulyako2014JRNC,planteur2012PC,roberts2003RA}.

\begin{longtable*}{|c|c|c|c|}
\caption{GROWTH METHODS}\\
\hline
\label{table1}
\textbf{Compound	}  & \textbf{					Synthesis Method                                                                                                             }  & \textbf{Form	} &\textbf{		References}  \\
\hline
\endfirsthead
\multicolumn{4}{c}%
{\tablename\ \thetable\ -- \textit{Continued from previous page}} \\
\hline
\textbf{	Compound}  & \textbf{						Synthesis Method							}  & \textbf{	Form} &\textbf{		References} \\
\hline
\endhead
\endfoot
\hline
\endlastfoot
ThO$_2$ & Calcination of oxalate - air  & Powder & \cite{kurina2002AE,qi2022IJACT} \\
\hline 
 & Sol-gel & Powder & \cite{matthews1978thoria,wymer1965preparation} \\
\hline 
 & From melt & Single Crystal & \cite{herrick1981JCG} \\
\hline 
 & Flux - Lead and other flux& Single Crystal & \cite{finch1965JAP,wanklyn1984JCG,andreetti1985JCG,garton1972JCG,rizzoli1984ICA,linares1967JPCS,chase1964AMJEPM} \\
\hline 
 & Hydrothermal & Single Crystal & \cite{mann2010CGD,castilow2013MRSOPL} \\
\hline 
 & Sublimation/deposition& Single Crystal & \cite{laszlo1967JPCS} \\
\hline 
 & Electrodeposition from plasma & Single Crystal & \cite{kawabuchi1979JAP,kawabuchi1980JCG} \\
\hline 
PaO$_2$ & Reduction of oxide hydrate - H$_2$ & Single Crystal & \cite{sellers1954JACS} \\
\hline 
Pa$_2$O$_5$ & Calcination of various Pa compounds - air/O$_2$ & Powder & \cite{stchouzkoy1964CRHSAS} \\
\hline 
 & Dehydration of oxide hydrate & Powder & \cite{sellers1954JACS,kirby1961JINC} \\
\hline 
 & Precipitation of molten salt - HF/H$_2$O purge & Powder & \cite{tallent1974ORNL} \\
\hline 
UO$_2$ & Sol-gel  & Powder & \cite{wymer1965preparation} \\
\hline 
 & Reduction of U$_3$O$_8$ or UO$_3$ & Powder & \cite{belle1961uranium} \\
\hline 
 & Aqueous & Powder & \cite{kulyako2014JRNC} \\
\hline 
 & From melt & Single Crystal & \cite{herrick1981JCG,nasu1964JJAP,sakurai1968JCG} \\
\hline 
 & Floating Zone & Single Crystal & \cite{chapman1965JACS} \\
\hline 
 & Flux growth - UCl$_4$ & Single Crystal & \cite{schlechter1968JNM} \\
\hline 
 & Electrodeposition - chloride melt & Single Crystal & \cite{robins1961JNM,li2022JES,amelinckx1963growth} \\
\hline 
 & Hydrothermal & Single Crystal & \cite{rickert2021CGD,young2016PRL,dugan2018EPJB} \\
\hline 
 & Vapor Transport - chloride species or others & Single Crystal & \cite{naito1971JCG,faile1978JCG,singh1974JCG,spirlet1979JPC} \\
\hline 
 & Sublimation & Single Crystal & \cite{amelinckx1963growth,van1962JNM} \\
\hline 
 & Electrodeposition from plasma & Single Crystal & \cite{kawabuchi1980JCG} \\
\hline 
U$_4$O$_9$ & Comproportionation of UO$_2$ and U$_3$O$_8$ & Powder & \cite{soulie2019IC,osborne1957JACS} \\
\hline 
 & Oxidation of UO$_2$ & Single Crystal or Powder & \cite{masaki1972ACS,leinders2020IC,desgranges2009IC,masaki1970JCG,hoekstra1961JINC,bevan1986JSSC} \\
\hline 
 & Electrodeposition - chloride melt & Single Crystal & \cite{robins1961JNM} \\
\hline 
 & Vapor transport - HCl or Cl$_2$ & Single Crystal & \cite{naito1971JCG} \\
\hline 
U$_3$O$_7$  & Oxidation of UO$_2$ & Powder & \cite{leinders2020IC} \\
\hline 
$\beta$-U$_3$O$_7$ & Oxidation of UO$_2$ & Powder & \cite{desgranges2009IC,hoekstra1961JINC,westrum1962JPCS,allen1986JCSFT,garrido2003JNM} \\
\hline 
U$_2$O$_5$ & Comproportionation of UO$_2$ and U$_3$O$_8$ - High P & Powder & \cite{hoekstra1970JINC} \\
 \hline 
 & Reduction of U$_3$O$-8$ single crystals & Single Crystal & \cite{spitsyn1972DANSSSR} \\
\hline
 & Decomposition of UO$_2$Cl$_2$& Single Crystal & \cite{rundle1948JACS} \\
\hline
$\alpha$-U$_3$O$_8$ & Calcination of oxalate or nitrate - air/He/O$_2$ & Powder & \cite{desfougeres2020IC,vita1973ACA}  \\
\hline 
 & UO$_2$ Oxidation & Powder & \cite{song2002JNST,loopstra1970JAC} \\
\hline 
 & UO$_3$ Reduction & Powder & \cite{zhang2014JSSC} \\
\hline 
 & Vapor Transport - HCL & Single Crystal & \cite{naito1971JCG} \\
\hline 
 & UO$_2$Cl$_2$ decomposition & Single Crystal &\cite{robins1962JNM}  \\
\hline 
$\beta$-U$_3$O$_8$ & Heat treating $\alpha$-U$_3$O$_8$ & Powder & \cite{loopstra1970ACB} \\
\hline 
 & Sublimation/Deposition & Single Crystal &\cite{hoekstra1955JPC}  \\
\hline 
$\alpha$-UO$_3$ & Crystallization of amorphous UO$_3$ - High pressure O$_2$ & Powder & \cite{robins1961JNM} \\
\hline
 & Dehydration of H$_2$U$_3$O$_{10}$ & Powder &\cite{wheeler1964JINC}  \\
\hline
 & Calcination of UO$_4$2H$_2$O or NH$_4$UO$_2$(NO$_3$)$_3$ - air or others& Powder &\cite{sweet2013JRNC,greaves1972ACB,kim2012JRNC,cordfunke1961JINC}  \\
\hline
 &Oxidation of U$_3$O$_8$ - High pressure O$_2$ & Powder &\cite{sheft1950JACS}  \\
\hline
$\beta$-UO$_3$ & Calcination of nitrate or NH$_4$UO$_2$(NO$_3$)$_3$ - air or others & Powder &\cite{kim2012JRNC,debets1966AC,spano2020IC}  \\
\hline
 & Oxidation of U$_3$O$_8$ & Powder &\cite{sheft1950JACS}  \\
\hline
 $\gamma$-UO$_3$ & Calcination or nitrate or (NH$_4$)$_2$UO$_2$(NO$_3$)$_4$2H$_2$O - air or others & Powder &\cite{kim2012JRNC,sweet2013JRNC,engmann1963AC}  \\
\hline
 & Flux - ZnCl$_2$ & Single Crystal &\cite{hoekstra1966JINC}  \\
\hline
 $\delta$-UO$_3$ & Dehydration of $\beta$-UO$_3$H$_2$O& Powder &\cite{wait1955JINC,hoekstra1961JINC,weller1988Poly}  \\
\hline
 $\epsilon$-UO$_3$ & Oxidation of U$_3$O$_8$ - NO$_2$/O$_2$/O$_3$ & Powder &\cite{hoekstra1961JINC,katz1949JACS,spano2022JNM}  \\
\hline
UO$_4$ & Aqueous & Powder &\cite{planteur2012PC}  \\
\hline
NpO$_2$ & Calcination of oxalate or other compounds & Powder &\cite{katz1949JACS,yarbro1991LANL,bessonov1989R,richter1987JNM,fahey1974INCL,bagnall1964JCS}  \\
\hline 
 & Aqueous & Powder &\cite{kulyako2014JRNC,roberts2003RA}  \\
\hline
 & Flux - Lead or LiMoO$_2$ & Single Crystal &\cite{rizzoli1984ICA,finch1970JCG}  \\
\hline
 & Electrochemical - Cl melt  & Single Crystal &\cite{martinot1970BSCB}  \\
\hline 
 & Vapor transport - TeCl$_4$ & Single Crystal &\cite{spirlet1979JPC,aoki2006JPSJ,spirlet1980JCG}  \\
\hline  
Np$_2$O$_5$ & Calcination of various Np compounds - air/vacuum& Powder &\cite{bessonov1989R,bagnall1964JCS,merli1994RA}  \\
\hline 
 & Flux - LiClO$_4$ & Powder &\cite{cohen1964JACS,cohen1963IC}  \\
\hline
 & Aqueous & Powder &\cite{nitsche1993LANL}  \\
\hline
 & Mild hydrothermal & Single Crystal &\cite{forbes2007JACS}  \\
\hline
Pu$_2$O$_3$ & Reduction of PuO$_2$ - H$_2$ or C & Powder &\cite{flotow1981JCP,chikalla1964JNM}  \\
\hline
PuO$_2$ & Calcination of various Pu compounds - air & Powder &\cite{bagnall1964JCS,drummond1957JACS}  \\
\hline
 & Oxygenation of PuF$-4$ - H$_2$O & Powder &\cite{flotow1976JCP}  \\
\hline
 & Sol-gel & Powder &\cite{wymer1965preparation,plymale1967preparation,lloyd1968NA}  \\
\hline
 & Aqueous & Powder &\cite{kulyako2014JRNC}  \\
\hline
 & Flux -various & Single Crystal &\cite{rizzoli1984ICA,modin2008RSI,finch1972JCG,schlechter1970JNM}  \\
\hline
 & Vapor transport - TeCl$_4$ & Single Crystal &\cite{rebizant2000AIPCP}  \\
\hline
 & Precipitation from silicate glass & Single Crystal &\cite{phipps1964S}  \\
\hline
Am$_2$O$_3$ & Reduction of AmO$_2$ - H$_2$ & Powder &\cite{morss1985JNM,eyring1952JACS,hurtgen1977INCL}  \\
\hline
AmO$_2$ & Calcination of oxalate - air/O$_2$ & Powder &\cite{fahey1974INCL,morss1985JNM,eyring1952JACS,baybarz1960ORNL,morss1981JINC}  \\
\hline
 & Calcination of exchange resin - O$_2 $ & Powder &\cite{hurtgen1977INCL}  \\
\hline
Cm$_2$O$_3$ & Calcination of oxalate - CO/vacuum & Powder &\cite{nave1983PRB}  \\
\hline
 & Calcination of exchange resin - air  & Powder &\cite{noe1970INCL}  \\
\hline
 & Reduction of CmO$_2$ - H$_2$/Ar$_2$/vacuum & Powder &\cite{morss1983IC,asprey1955JACS,posey1973ORNL}  \\
\hline
CmO$_2$ & Calcination of oxalate or nitrate - O$_2$ or O$_2$-O$_3$ mixture & Powder &\cite{nave1983PRB,asprey1955JACS,haug1967JINC}  \\
\hline
 & Calcination of exchange resin - O$_2 $ & Powder &\cite{noe1971INCL,hale1973JINC}  \\
\hline
 & Oxidation of Cm$_2$O$_3$ - O$_2$& Powder &\cite{peterson1971JINC}  \\
\hline
BkO$_2$ & Calcination of oxalate - air/O$_2$ & Powder &\cite{fahey1974INCL,nave1983PRB,turcotte1980JINC}  \\
\hline
 & Calcination of exchange resin & Powder &\cite{peterson1967INCL,baybarz1968JINC}  \\
\hline
Cf$_2$O$_3$ & Calcination of exchange resin - air & Powder &\cite{baybarz1972JINC}  \\
\hline
 & Reduction of CfO$_2$ - H$_2$ & Powder &\cite{morss1987JLCM}  \\
\hline
 CfO$_2$ & Calcination of exchange resin - O$_2$ & Powder &\cite{baybarz1972JINC}  \\
\hline
\end{longtable*}

Many of the actinide oxides can be hypo- or hyper-stoichiometric, i.e. containing less or more oxygen than the integer stoichiometry would imply. These require specific synthesis conditions or post-synthetic treatments to achieve stoichiometric products. Hypo- or hyper-stoichiometry has been extensively studied in the uranium oxides. For example, highly variable stoichiometry exists between UO$_2$ and U$_3$O$_8$. This includes several distinct stoichiometries, such as U$_4$O$_9$, U$_3$O$_7$, and U$_2$O$_5$.  Careful control of the oxidation of UO$_2$\cite{leinders2020IC,masaki1970JCG,soulie2019IC,hoekstra1961JINC} or reduction of U$_3$O$_8$\cite{spitsyn1972DANSSSR} allows for the synthesis of these various compounds. For example, U$_4$O$_9$ can be synthesized by the comproportionation of UO$_2$ and U$_3$O$_8$\cite{soulie2019IC}. It can also be synthesized by the oxidation of powder or single crystalline samples of UO$_2$ using either a low oxygen concentration\cite{leinders2020IC} or an adjacent U$_3$O$_8$ sample, which releases oxygen upon heating, in a sealed system\cite{masaki1970JCG}. McEachern and Taylor provided a thorough review of the oxidation behavior of uranium dioxide\cite{mceachern1998JNM}.

In contrast to polycrystalline samples, single crystals offer the highest-quality specimens and enable the measurement of anisotropic properties, which are crucial for determining their often intricate magnetic characteristics. Techniques such as neutron scattering, resonant X-ray scattering, electronic and heat transport, and specific heat measurements become invaluable tools in this regard. Actinide materials exhibit complex magnetic behavior due to strong electronic correlations, spin-orbit coupling, and crystal-field interactions, underscoring the importance of precise and comprehensive characterization. However, achieving single crystal growth of actinide oxides presents an added layer of complexity in material preparation and handling, especially when compared to non-radioactive substances. For instance, when working with reactive, radioactive, and toxic elements, it is typically necessary to seal the materials in a quartz tube during crystal growth. One notable advantage of single crystals is that, in many instances, they require a significantly smaller quantity of material compared to polycrystalline samples for magnetic characterization, such as magnetic susceptibility and specific heat measurements, as well as neutron diffraction. Moreover, single crystals are the preferred choice when working with actinide materials for measurements, as polycrystalline powder materials have a higher potential for contamination.

The crystal growth of actinide oxides is made even more challenging by their very high melting points, their proclivity for auto-reduction at high temperatures, and their high inertness, especially of the dioxides. For instance, ThO$_2$ has the highest melting point of any known binary oxides reaching up to 3,300 $^{\circ}$ \cite{emsley2001nature}. Despite their high melting points, single crystals of ThO$_2$ and UO$_2$ have been synthesized from their melts using synthesis methods that allow for extremely high temperatures, such as solar furnaces\cite{sakurai1968JCG}, electric arc furnaces\cite{brite1961ORNL}, and RF furnaces\cite{herrick1981JCG}. Single crystals of UO$_2$ have also been achieved using a floating zone furnace\cite{chapman1965JACS}. 

The difficulties of direct melt growth of actinide oxides make crystal growth techniques that can provide single crystal samples at more moderate temperatures very desirable\cite{juillerat2019DT,klepov2020FC,wang2021JSSC}.  Flux crystal growth\cite{bugaris2012ACIE} has been reported for AcO$_2$ (Ac = Th, U, Np, Pu). Lead containing fluxes, namely PbF$_2$, PbO, and lead vanadate\cite{rizzoli1984ICA,wanklyn1984JCG} are the most commonly used, but other fluxes including lithium tungstate, NaBO$_2$, and MoO$_2$ have also been reported\cite{finch1965JAP,modin2008RSI,linares1967JPCS}. In addition to flux growth, where precipitation is typically driven by cooling or evaporation of the flux, electrochemical crystal growth from chloride melts has been reported for UO$_2$ and NpO$_2$\cite{martinot1970BSCB,li2022JES}. 

Along with using molten salts as a solvent, recent years have seen a number of developments in the use of hydrothermal synthesis wherein supercritical water is the solvent\cite{forbes2007JACS,mcmillen2012PM} to grow crystals of ThO$_2$ and UO$_2$\cite{mann2010CGD,rickert2021CGD}. For example, large single crystals of ThO$_2$, up to several $mm$ per side, were grown across a temperature gradient from seed crystals at 710 $^{\circ}$C using polycrystalline ThO$_2$ as a source material in 6 M CsF at 750 $^{\circ}$C\cite{mann2010CGD}. Mild hydrothermal synthesis\cite{popa2018CEC}, which occurs in the temperature range above the boiling point of water but below the supercritical point, and aqueous routes\cite{kulyako2014JRNC,husar2015CC}, which occur below the boiling point of water, have also been reported for some of the actinide oxides, but have only resulted in polycrystalline samples\cite{kulyako2014JRNC} or nanoparticles\cite{popa2018CEC,husar2015CC}.

A final, commonly reported growth method for the synthesis of single crystals is vapor transport\cite{binnewies2013ZAAC}, a method in which a polycrystalline material or precursor mixture is transported across a temperature gradient by a transport gas and redeposited as single crystals. Vapor transport has been reported for NpO$_2$, PuO$_2$ and several of the uranium oxides\cite{naito1971JCG,spirlet1979JPC,rebizant2000AIPCP}. A chloride containing species, namely HCl, Cl$_2$, or TeCl$_4$, is the most common transport agent. ThO$_2$ has been grown by the direct sublimation/deposition of polycrystalline ThO$_2$ without the use of a transport gas\cite{laszlo1967JPCS}. A graphical representation of synthesis methods is presented in Fig. \ref{Graphics}.

As will be discussed in the Magnetic Properties section, achieving a better understanding of the behavior of many of the binary actinide oxides necessitates further measurements, especially on single crystalline samples.  The provided summary and Table \ref{table1} can serve as a guide for synthesizing such samples. For binary actinide oxides with only a limited number of synthesis procedures, especially where only polycrystalline products are obtained, the synthesis methods of other actinides, with proper consideration of oxidation states, can guide the development of suitable routes for synthesis. For later actinides particularly, issues both of supply and high radioactivity limit the amount of material available, complicating synthesis. For such species, hydrothermal synthesis has been shown to be scalable down to very small sample reaction masses,\cite{cross2012IC,deason2022JACS,cary2015NC,sykora2004JSSC} using as little as $<$ 1 mg of actinide\cite{sykora2004JSSC}. Recently, we have shown that flux growth can also be scaled down to the mg scale or even 1 mg scale\cite{deason2023JACS,pace2020CC}. This suggests that these two synthesis techniques are likely suitable for the synthesis of many of the binary actinide oxides whose complete magnetic characterization necessitates single crystalline samples.


\begin{longtable*}{|l|c|c|c|c|c|c|c|}
\caption{Crystal structure information of Actinide Oxides} \\
\hline
\label{table2}
\textbf{Compound} & \textbf{Crystal Symmetry} & \textbf{Space Group} & \multicolumn{4}{|c|}{\textbf{Lattice Parameters(\AA)}} & \textbf{References} \\
{} & {} & {} & \textbf{a} & \textbf{b} & \textbf{c} & \textbf{$\beta$} & {} \\
\hline
\endfirsthead
\multicolumn{8}{c}%
{\tablename\ \thetable\ -- \textit{Continued from previous page}} \\
\hline
\textbf{Compound} & \textbf{Crystal Symmetry} & \textbf{Space Group} & \multicolumn{4}{|c|}{\textbf{Lattice Parameters(\AA)}} & \textbf{References} \\
{} & {} & {} & \textbf{a} & \textbf{b} & \textbf{c} & \textbf{$\beta$} & {} \\
\hline
\endhead
\hline
\endfoot
\hline
\endlastfoot

ThO$_2$ & Cubic & Fm$\overline{3}$m & 5.597 & {} & {} & {} & \cite{Yamashita1997jnucmat} \\
PaO$_2$ & Cubic & Fm$\overline{3}$m & 5.505 & {} & {} & {} & \cite{sellers1954JACS} \\
Pa$_2$O$_3$ & Cubic & Fm$\overline{3}$m & 5.505 & {} & {} & {} & \cite{sellers1954JACS} \\
NpO$_2$ & Cubic & Fm$\overline{3}$m & 5.434 & {} & {} & {} & \cite{Yamashita1997jnucmat} \\
Np$_2$O$_5$ & Monoclinic & P2/c & 8.168 & 6.584 & 9.313 & 116.09 & \cite{forbes2007JACS} \\
PuO$_2$  & Cubic & Fm$\overline{3}$m & 5.395 & {} & {} & {} &\cite{Yamashita1997jnucmat}  \\
$\alpha$-Pu$_2$O$_3$ & Cubic & Ia\={3} & 11.04 & {} & {} & {} & \cite{gardner1965JINC} \\
$\alpha$$^\prime$-Pu$_2$O$_3$ & Cubic & Ia\={3} & 5.409 & {} & {} & {} & {\cite{gardner1965JINC}} \\
$\beta$-Pu$_2$O$_3$ & Hexagonal & \textit{P}6$_3$/mmc & 3.841 & 3.841 & 5.958 & {} & \cite{gardner1965JINC} \\
AmO$_2$ & Cubic & Fm$\overline{3}$m & 5.139 & {} & {} & {} & \cite{Dancausse2002jnst} \\
Am$_2$O$_3$ & Monoclinic & C2/m & 14.315 & 3.6793 & 8.9271 & 100.37 & \cite{Horlait2014JSSC,chikalla1968JINC} \\
{} & Hexagonal & P$\overline{3}$m1 & 3.8123 & 3.8123 & 5.9845 & {} & \cite{hurtgen1977INCL,Horlait2014JSSC,chikalla1968JINC} \\
{} & Cubic & Ia$\overline{3}$ & 11.012 & {} & {} & {} & \cite{hurtgen1977INCL,Horlait2014JSSC,chikalla1968JINC} \\
CmO$_2$ & Cubic & Fm$\overline{3}$m & 5.358 & {} & {} & {} & \cite{nave1983PRB,morss1989JLCM,Dancausse2002jnst} \\
Cm$_2$O$_3$ & Monoclinic & C2/m & 14.282 & 3.652 & 8.900 & 100.31 & \cite{nave1983PRB,Mosley1972jinc} \\
{} & Hexagonal & P$\overline{3}$m1 & 3.80 & 3.80 & 6.00 & {} & \cite{nave1983PRB} \\
{} & Cubic & Ia$\overline{3}$ & 10.97 & {} & {} & {} & \cite{wallmann1964JINC} \\
BkO$_2$  & Cubic & Fm$\overline{3}$m & 5.333 & {} & {} & {} &\cite{nave1983PRB,Peterson1967inclet}  \\
Bk$_2$O$_3$ & Cubic & \textit{I}a$\overline{3}$ & 10.887 & {} & {} & {} & \cite{Peterson1967inclet} \\
{} & Monoclinic & C2/m & 14.197 & 3.606 & 8.846 & 100.23 & \cite{Baybarz1972aUSatom} \\ 
CfO$_2$ & Cubic & Fm$\overline{3}$m & 5.310 & {} & {} & {} & \cite{baybarz1972JINC} \\
Cf$_2$O$_3$ & Cubic & Fm$\overline{3}$m & 10.809 & {} & {} & {} & \cite{baybarz1972JINC, copeland1969JINC,Copeland1967UCBerk} \\
{} & Monoclinic & C2/m & 14.124 & 3.591 & 8.809 & 100.31 & \cite{copeland1969JINC,green1965UCB} \\

\hline
\multicolumn{8}{|c|}{* =  Magnetic properties and the magnetic ground state have not been studied in detail or remain unresolved} \\
\hline

\end{longtable*}

\begin{longtable*}{|l|c|c|c|c|c|c|c|}
\caption{Crystal structure information of Uranium Oxides} \\
\hline
\label{table3}
\textbf{Compound} & \textbf{Crystal Symmetry} & \textbf{Space Group} & \multicolumn{4}{|c|}{\textbf{Lattice Parameters(\AA)}} & \textbf{References} \\
{} & {} & {} & \textbf{a} & \textbf{b} & \textbf{c} & \textbf{$\beta$} & {} \\
\hline
\endfirsthead
\multicolumn{8}{c}%
{\tablename\ \thetable\ -- \textit{Continued from previous page}} \\
\hline
\textbf{Compound} & \textbf{Crystal Symmetry} & \textbf{Space Group} & \multicolumn{4}{|c|}{\textbf{Lattice Parameters(\AA)}} & \textbf{References} \\
{} & {} & {} & \textbf{a} & \textbf{b} & \textbf{c} & \textbf{$\beta$} & {} \\
\hline
\endhead
\hline
\endfoot
\hline
\endlastfoot

UO$_2$ & Cubic & Fm$\overline{3}$m & 5.47 & 5.47 & 5.47 & {} &\cite{lambertson1953JACerS,LEINDERS2015jnucmat,Yamashita1997jnucmat}  \\
$\alpha$-UO$_3$ & Orthorhombic & C2mm & 3.913 & 6.936 & 4.167 & {} & \cite{loopstra1966RTCPB,enriquez2020ACSAMI} \\ 
{} & {} & C222 & 6.84 & 43.45 & 4.157 & {} & \cite{greaves1972ACB} \\
$\beta$-UO$_3$ & Monoclinic & P2$_1$/m & 10.34 & 14.33 & 3.910 & {99.03} & \cite{enriquez2020ACSAMI,debets1966AC} \\
$\gamma$-UO$_3$ & Orthorhombic & F$_\textit{ddd}$ & 9.79 & 19.93 & 9.71 & {} & \cite{loopstra1977JSSC} \\
{} & Tetragonal & I4$_1$ & 6.90 & 6.90 & 19.98 & {} & \cite{loopstra1977JSSC} \\
$\delta$-UO$_3$ & Cubic & Pm$\overline{3}$m & 4.165 & 4.165 & 4.165 & {} & \cite{weller1988Poly} \\
$\eta$-UO$_3$ & Orthorhombic & P2$_1$2$_1$2$_1$ & 7.51 & 5.47 & 5.22 & {} & \cite{siegel1966AC} \\
$\alpha$-U$_3$O$_8$ & Orthorhombic & Amm2 & 6.716 & 11.960 & 4.147 &  & \cite{loopstra1966RTCPB,loopstra1970JAC,ranasinghe2020CMS} \\
{} & Hexagonal & P$\overline{6}$2m & 6.812 & 6.812 & 4.142 & 120 & \cite{loopstra1970JAC} \\
$\beta$-U$_3$O$_8$ & Orthorhombic & Cmcm & 7.069 & 11.445 & 8.303 & {} & \cite{loopstra1970JAC,loopstra1970ACB} \\
U$_3$O$_7$ & Cubic & P4$_2$/nnm & 5.3799 & 5.3799 & 5.5491 & {} & \cite{leinders2016IC} \\
$\alpha$-U$_2$O$_5$ & Unk. & Unk. & Unk. & Unk. & Unk. & Unk. & \cite{hoekstra1970JINC,kovba1979R} \\
$\beta$-U$_2$O$_5$ & Hexagonal & Unk. & 3.813 & 3.813 & 13.180 & {} & \cite{hoekstra1970JINC} \\
$\gamma$-U$_2$O$_5$ & Monoclinic & Unk. & 5.410 & 5.481 & 5.410 & 90.49 & \cite{hoekstra1970JINC} \\
$\delta$-U$_2$O$_5$ & Orthorhombic & Pnma & 6.849 & 8.274 & 31.706 & {} & \cite{kovba1979R} \\
$\alpha$-U$_4$O$_9$ & Cubic & R3c & 18.9286 & 18.9286 & 18.9286 & {} & \cite{desgranges2011IC,desgranges2009IC} \\
$\beta$-U$_4$O$_9$ & Cubic & I$\overline{4}$3d & 21.766 & 21.766 & 21.766 & {} & \cite{desgranges2009IC} \\
$\gamma$-U$_4$O$_9$ & Cubic & I$\overline{4}$3d & 21.944 & 21.944 & 21.944 & {} & \cite{desgranges2016IC} \\

\hline
\multicolumn{8}{|c|} {Note: The magnetic properties and magnetic ground state of the above uranium oxides, except for UO$_2$ and U$_3$O$_8$, have not }\\
\multicolumn{8}{|c|} {been studied in detail or remain unresolved} \\
\hline

\end{longtable*}

\section{Magnetic properties}\label{sec4}

\begin{figure}[t!]
\includegraphics[clip,width=\columnwidth]{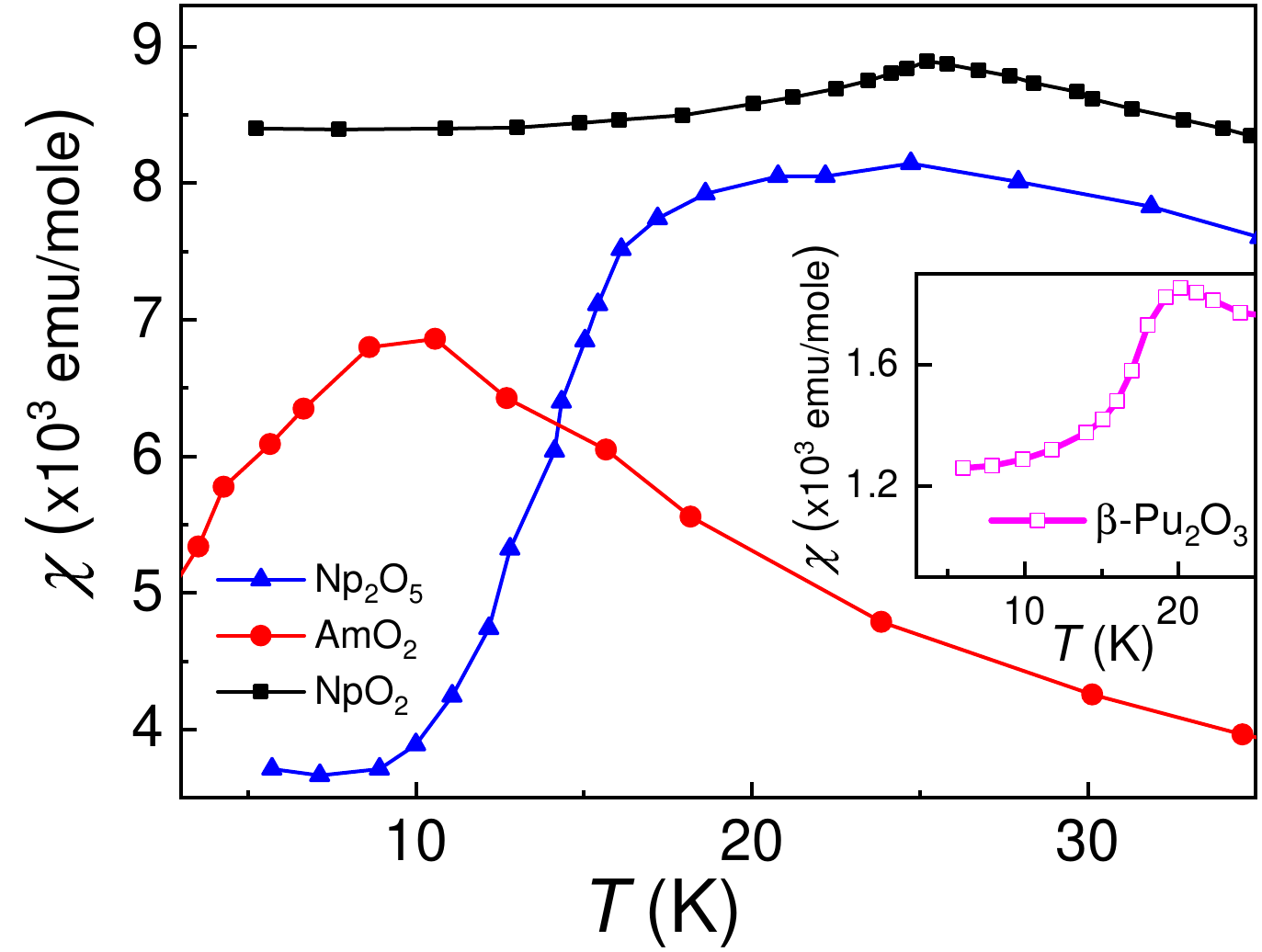}
\caption{\label{An_sus} Magnetic susceptibility of  NpO$_2$\cite{erdos1980PBC},   AmO$_2$\cite{karraker1975JCP_B}, $\beta$-Pu$_2$O$_3$\cite{mccart1981JCP} and Np$_2$O$_5$\cite{forbes2007JACS} as a function of temperature showing a cusp anomaly related to magnetic phase transitions.}
\end{figure}

The magnetism of binary actinide oxides through californium are discussed. The oxidation states of these actinides within these compounds are complex. If these actinide compounds follow the same magnetic systematics as the lanthanide compounds and, assuming the ground state is defined by Hund's-rule and $L-S$ coupling ($L$ being the total orbital angular momentum, $S$ being the total spin angular momentum), the ionic-like compounds which correspond to 5$f$ electronic configurations should have finite paramagnetic effective moments, with the exceptions of  the 5$f^4$ (singlet ground state) and 5$f^6$ ($J$=0) states that lead to a theoretical moment of 0 $\mu_B$. For example, Am$_2$O$_3$ and CmO$_2$ should be ionic-like compounds corresponding to the 5$f^6$ electronic configuration and therefore, should have theoretical values of 0 $\mu_B$ for their paramagnetic effective moment. Cm$_2$O$_3$ and BkO$_2$ should be ionic-like compounds corresponding to the 5$f^7$ configuration, with $J = S =\frac{7}{2}$ ($J$ being the total angular momentum), and an expected effective moment of 7.94  $\mu_B$. However, $L-S$ coupling may not be the most appropriate scheme for describing the ground-state wave functions in heavy actinides due to the large spin-orbit coupling and significant mixing of Hund’s rule ground states by other $LS$ states with the same $J$ value. Thus, an intermediate coupling model between $L-S$ and $j-$j coupling may be necessary\cite{nave1983PRB}. The magnetic ordering temperatures and effective moments of actinide oxides through californium are shown in Fig.\:\ref{TN,Mu}. Magnetic susceptibilities of various binary actinide oxides are presented in Fig.\:\ref{An_sus} and \ref{U_sus}.

In general, in the 5$f$-electron materials with strong spin-orbit coupling, dipole moments (the first term of a multipolar expansion) are insufficient to explain the observed phase transitions or quantum phenomena and, therefore, higher terms of the multipolar expansion are required. For example, multipolar ordering in NpO$_2$ arises at the atomic scale from strong spin-orbit coupling, which is difficult to detect because of the lack of response of high-rank multipoles to external perturbations. In other words, the characterization of the anisotropic distributions of electric and magnetic charges around given points of the crystal structure is challenging. Similar to NpO$_2$, M\"{o}ssbauer spectroscopy\cite{kalvius1969PLB} and neutron diffraction measurements failed to detect any magnetic order below $T_0$ = 8.5 K in AmO$_2$\cite{boeuf1979JDPL}. However, NMR measurements on AmO$_2$ performed by $Tokunaga$ $et$ $al.$\cite{tokunaga2014PRB,tokunaga2010JPSJ} confirmed the phase transition seen at $T_0$, although the NMR spectrum of AmO$_2$ is rather different from that seen in the antiferromagnetic ground state of UO$_2$\cite{ikushima2001PRB} or the multipolar ground state of NpO$_2$\cite{tokunaga2011JPSJ}.  

For some actinide oxides, their true magnetic ground state remains undetermined despite prior investigation of their magnetic properties. For example, PuO$_2$ shows a temperature independent magnetic susceptibility in between $4-1000$ K \cite{raphael1968SSC} and NMR measurements have confirmed the absence of any magnetic order or structural distortions down to $6$ K\cite{tokunaga2007JAC}. However, several theoretical studies\cite{sun2008JCP,jomard2008PRB,pegg2018PCCP,pegg2018JPCC} have predicted an  antiferromagnetic (AFM) ground state for PuO$_2$. Based on the ionic picture, CmO$_2$ should have a singlet, nonmagnetic ground state and the magnetic susceptibility should behave as in PuO$_2$. In actuality, the magnetic susceptibility of CmO$_2$ is temperature-dependent, shows a large effective moment of 3.36(6) $\mu_B$\cite{nave1983PRB,morss1989JLCM}, and is suggested to have a mixed valence configuration of 5$f$ electrons\cite{huang2020PRB}.

While the antiferromagnetic structure in UO$_2$ (fluorite type, Fm$\bar{3}$m s.g. crystal structure) was first reported by Frazer et al. in 1965\cite{frazer1965PR}, the internal distortion of oxygen atoms within the ordered state was discovered using high-quality single crystals in 1975\cite{ faber1975PRL}. More recently, it was shown that the dynamic Jahn-Teller distortion\cite{ caciuffo1999PRB} (internal distortion) causes change in the space group from Fm$\bar{3}$m to Pa$\bar{3}$\cite{dorado2010PRB, santini2009RMP}. This detailed crystal structure information, obtained from single crystals of UO$_2$, has proven invaluable in comprehending both the complex magnetic structure and phonon dispersion. Such in-depth studies of other actinide oxides, particularly higher actinide oxides, have seldom (if ever) been explored experimentally, and their magnetic ground states often remain unresolved. For instance, no magnetic characterization, including thermodynamic measurements, has been conducted on oxides like U$_2$O$_5$,  U$_3$O$_7$  and Pa$_2$O$_5$. In the following sections, we provide a review of the current magnetic properties reported for actinide dioxides in Sec.\:\ref{sec4a}, the actinide sesquioxides in Sec.\:\ref{sec4b}, and various selected higher oxides in Sec.\:\ref{sec4c}.

\begin{figure}[t!]
\includegraphics[clip,width=\columnwidth]{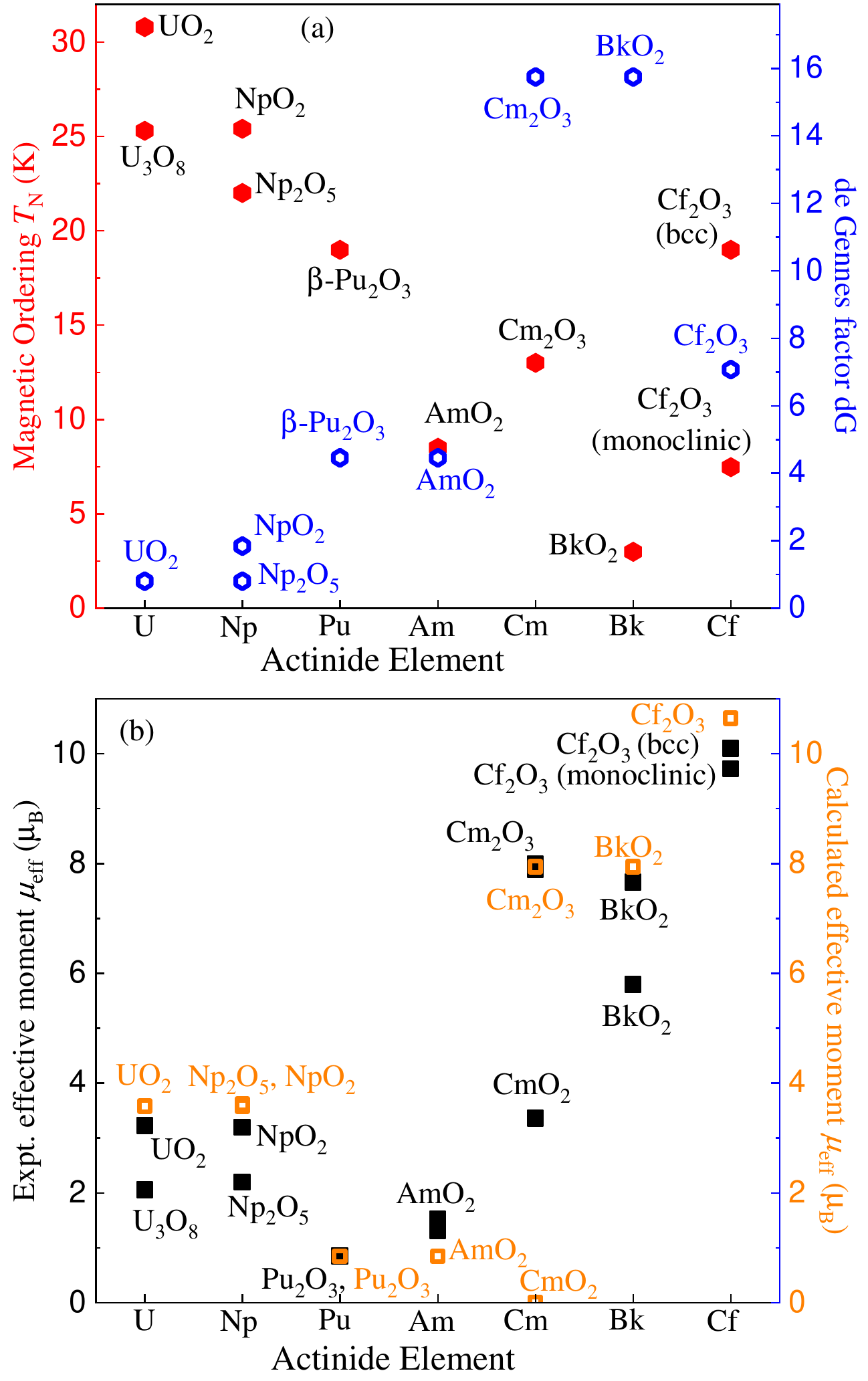}
\caption{\label{TN,Mu} (a) Magnetic ordering temperatures (left-axis) and de Gennes factor dG (right-axis) (b) experimental effective moments (left-axis) and theoretical effective moment (right-axis) of actinide elements: UO$_2$ \cite{kubo2005JPSJ}, U$_3$O$_8$\cite{leask1963JCS,westrum1959JACS}, U$_4$O$_9$\cite{gotoo1965JPCS}, NpO$_2$\cite{erdos1980PBC}, Np$_2$O$_5$\cite{forbes2007JACS}, $\beta$-Pu$_2$O$_3$\cite{mccart1981JCP}, AmO$_2$\cite{karraker1975JCP}, CmO$_2$\cite{nave1983PRB,morss1989JLCM}, Cm$_2$O$_3$\cite{morss1983IC}, BkO$_2$\cite{karraker1975JCP_B}, Cf$_2$O$_3$\cite{morss1987JLCM}.}
\end{figure}

 \subsection{Magnetic properties of actinide dioxides}\label{sec4a}

\subsubsection{Thorium Dioxide: ThO$_2$}
Due to the stability of ThO$_2$ and its importance in the nuclear technology, it is one of the most widely studied and well-characterized actinide dioxides. ThO$_2$ is a diamagnetic insulator marked by a negative magnetic susceptibility ($\chi = -16 \times 10^{-6}$ cm$^3$/mol \cite{lide2004CRC}) and a heat capacity that does not show any sign of phase transitions \cite{dennett2020AM}, as well as a band gap near $6$ eV \cite{rodine1971PRB,pantelides1974PRB}. Other physical properties have been well studied and are summarized by $Belle$ $et$ $al.$\cite{belle1984thorium} and recently by $Hurley$ $et$ $al$ \cite{hurley2021CR}. ThO$_2$ crystallizes in a fluorite-like face-centered cubic ($fcc$) structure ($Fm\overline{3}m$ Space Group \cite{rodriguez1981JNM}) with a lattice parameter of $5.597$ \AA\cite{Yamashita1997jnucmat}. It is worth mentioning that all actinide dioxides are isostructural, sharing the same fluorite $fcc$ structure as that of ThO$_2$ and therefore, ThO$_2$ is often used as a matrix for studying the unique properties of the rarer transuranic elements. Under the pressure of $40$ GPa, ThO$_2$ undergoes a structural phase transition like other flourite structures from cubic to a high-pressure orthorhombic structure ($Pnma$ Space group, known as cotunnite)\cite{dancausse1990HPR,jayaraman1988P,idiri2004PRB}.

\subsubsection{Uranium Dioxide: UO$_2$} 
UO$_2$ has been shown to be rather complex as compared to ThO$_2$ due to the presence of 5f-electrons in its electronic configuration. A Mott-Hubbard insulator UO$_2$ crystallizes in the shared fluorite cubic structure seen in ThO$_2$ with a lattice parameter of $5.4731(4)$ \AA~and also displays the cotunnite orthorhombic structure at high pressure\cite{lambertson1953JACerS,idiri2004PRB,benedict1982JPL,singh2022SSS,wang2013PRB}.  In UO$_2$, uranium exhibits the $4+$ oxidation state, yielding a 5$f^2$ configuration based on the ionic picture. An initial specific heat experiment hinted at a magnetic phase transition\cite{jones1952JCP}. Shortly thereafter, neutron diffraction studies on UO$_2$ single crystals determined a first-order AFM phase transition at $T_N$ = 30.8 K\cite{frazer1965PR,willis1965PL}. After decades of research, the magnetic state below $T_N$ was determined to be characterized by an antiferromagnetic non-collinear 3-$k$ structure\cite{burlet1986JLCM,amoretti1989PRB,blackburn2005PRB,caciuffo2007JMMM,giannozzi1987JMMM} accompanied by Jahn-Teller (JT) distortions of the oxygen atoms in the $fcc$ fluorite structure\cite{allen1968PRa,allen1968PRb,faber1975PRL,faber1976PRB,santini2009RMP}. 

Inelastic neutron scattering experiments first reported a possible strong coupling between the magnons and the crystal lattice in UO$_2$\cite{dolling1966PRL,cowley1968PR}. The first-order nature of the phase transition at $T_N$ arises from couplings between magnetic dipoles and electric quadrupoles (3-$k$ type)\cite{santini2009RMP,carretta2010PRL,caciuffo2011PRB} as marked by a sharp transition in various thermodynamic measurements such as magnetic susceptibility\cite{jaime2017NaturComm}, heat capacity\cite{bryan2019PRM}, thermal expansion, and elastic constants \cite{walstedt2014CRP,martel2014IC,katz2007BOOK,allen1968PRb,jaime2017NaturComm}. Subsequently, a dynamic JT distortion model was proposed to explain the strong magnetoelastic interactions seen well above $T_N$\cite{amoretti1989PRB}. NMR studies\cite{ikushima2001PRB} support the idea of a 3-$k$ magnetic structure in UO$_2$ below $T_N$ and provide strong evidence for local distortions as well as an excitation spectrum that shows the presence of magnon–phonon coupling. Many first principle studies have further explained and successfully modeled the non-coplanar 3$k$ magnetic structure below $T_N$\cite{gryaznov2010PCCP,zhou2011PRB,zhou2022PRB,roy2008JCC,dorado2013JPCM}. Due to the non-collinear magnetic structure that breaks time-reversal symmetry in a non-trivial way seen below $T_N$ in UO$_2$, the existence of piezomagnetism is possible\cite{bar1985ZETF}. The piezomagnetic effect (together with trigonal distortion upon applied magnetic field along $111$ direction) has recently been observed in UO$_2$ crystals\cite{jaime2017NaturComm},  revealing broken time-reversal symmetry and strong coupling between uranium magnetic moments and lattice degrees of freedom.

\subsubsection{Neptunium Dioxide: NpO$_2$} 

NpO$_2$ crystallizes in the same $fcc$ fluorite structure as the other actinide oxides, with a lattice parameter of $5.434$\AA \cite{Yamashita1997jnucmat}. Based on the ionic picture, the 5f$^3$ configuration for Np$^{4+}$ ions in NpO$_2$ predicts the crystal field ground state to be a magnetic doublet, suggesting the smallest amount of magnetic exchange could result in subsequent magnetic ordering. Early magnetic susceptibility and specific heat measurements \cite{ross1967JAP} suggested a magnetic phase transition $T_0$ at $25$ K. Although NpO$_2$ is isostructural to UO$_2$ and the entropy calculated from the specific heat below the phase transition ($T_0$ = 25 K) is very similar to that of UO$_2$\cite{osborne1953JCP}, the magnetic ground state of NpO$_2$ turned out to be complex \cite{caciuffo1987SSC}. Despite the magnetic phase transition suggested by earlier experiments, low-temperature neutron diffraction\cite{cox1967JPCS,heaton1967JPCS,boeuf1983PSSAR} and M\"{o}ssbauer\cite{dunlap1968JPCS,friedt1985PRB} experiments have failed to identify an ordered magnetic moment. Fried et al.\cite{friedt1985PRB} later revisited the low-temperature M\"{o}ssbauer measurements down to 1.5 K with an applied magnetic field up to $8$ T suggesting no dipole magnetism and that the phase transition $T_0$ can be attributed to a hypothesized competitive structural transition. However, this structural transition, consisting of a JT distortion of the oxygen cage, has not been experimentally confirmed\cite{caciuffo1987SSC}.

To understand the phase transition $T_0$ of NpO$_2$ and address the experimental inconsistencies, Zolnierek et al.\cite{zolnierek1981JPCS} hypothesized the cation is trivalent in NpO$_2$. However, crystal field measurements using neutron spectroscopy displayed an inelastic peak originating from excitations between the $\Gamma{_8}^2$, and $\Gamma{_8}^1$ crystal field quartets and crystal field parameters are consistent with Np$^{4+}$ ions\cite{amoretti1992JPCM}. The observed ordering in NpO$_2$ with no net magnetic moment could be explained by quadrupolar ordering\cite{mannix1999PRB}. However, this cannot account for the observation of a temperature‐independent susceptibility\cite{erdos1980PBC} and the asymmetry in the muon experiments\cite{kalvius2001HPCRE}. However, octupolar ordering in NpO$_2$ could explain the essentially ‘zero’ moment of M\"{o}ssbauer spectroscopy and the asymmetry in the muon experiments. The possibility of having octupolar magnetic ordering in NpO$_2$ has been studied from a theoretical perspective\cite{shiina2007JPSJ,shiina2011JPSJ}. A resonant X-ray scattering study performed at the Np $M_{4,5}$ edges below $T_0$ has shown a growth of superlattice Bragg peaks related to the long-range order of Np electric quadrupoles\cite{mannix1999PRB,paixao2002PRL} and a complex ordering of higher-order magnetic multipoles, which lead to a singlet ground state with zero dipole-magnetic moment\cite{santini2009RMP, tokunaga2006PRL, suzuki2010PRB,walstedt2018PRB}. Unlike UO$_2$, no symmetry change have been observed in NpO$_2$, and recent neutron powder diffraction data collected down to 300 mK showed no significant changes in the diffraction pattern below and above $T_0$ = 25 K\cite{Frontzek2022}. Frontzek et al.\cite{Frontzek2022} suggested that neutron single crystal diffraction on NpO$_2$ could improve the detection limit for octupolar order. 


\subsubsection{Plutonium Dioxide: PuO$_2$} 

Despite PuO$_2$ being used in the nuclear fuel industry and being the subject of nearly $80$ years of research, its ground state is still a subject of extensive studies. PuO$_2$ also crystallizes in the same $fcc$ fluorite structure as the other actinide oxides, with a lattice parameter of $a$ = 5.395\AA \cite{Yamashita1997jnucmat}. Laser heating studies have shown PuO$_2$ melts at $3017(28)$ K\cite{de2011JNM}, several hundred degrees higher than the previously cited melting point of $~2700$ K\cite{mccleskey2013JAP}. Recent X-ray absorption near edge spectroscopy (XANES) and X-ray absorption fine structure (EXAFS) measurements suggest an optical band gap of 2.8(1) $eV$ at room temperature\cite{mccleskey2013JAP}. Furthermore, the low-temperature electronic contribution to the specific heat in PuO$_2$ can be extrapolated to zero at $T = 0$, in agreement with its Mott-insulating ground state \cite{hurley2021CR}. 

Based on the ionic picture, the Pu$^{4+}$ ion with a 5$f^4$ configuration in PuO$_2$ suggests the crystal field ground state to be a $\Gamma_1$ singlet\cite{magnani2005PRB,shick2014PRB}. The presence of a non-magnetic ground state in PuO$_2$ is supported by the specific heat of PuO$_2$ (polycrystalline samples) that shows no anomaly in the $C_p(T)$\cite{hurley2021CR}. Early experimental evidence pointed to a paramagnetic ground state with a constant magnetic susceptibility of $0.54$x$10^{-3}$ $cm^3$$mol^{-1}$ up to $1000$ K \cite{raphael1968SSC} suggesting Van Vleck paramagnetism due to the crystal electric field transition between the $\Gamma_1$ ground state and the $\Gamma_4$ state \cite{raphael1968SSC}. This understanding was confirmed through inelastic neutron scattering (INS) measurements; however, the transition was measured to be $123$ $meV$, suggesting a magnetic susceptibility of $\sim1$x$10^{-3}$ $cm^3$$mol^{-1}$\cite{kern1999PRB}. The constant magnetic susceptibility measured in between $4-1000$ K \cite{raphael1968SSC} estimated the transition energy for $\Gamma_1~\rightarrow ~\Gamma_4$ to be $\approx$ 284 meV, much higher than measured from the INS experiments. Subsequent NMR measurements confirmed the absence of any magnetic order or structural distortions down to $6$ K, verifying the temperature independence of the magnetic susceptibility\cite{tokunaga2007JAC}. 

It is worth mentioning that the magnetic ground state of PuO$_2$ is a subject of ongoing, mostly theoretical, studies that might suggest the existence of antiferromagnetic order in this system. A first-principles calculation of the crystal field scheme reported the transition energy for $\Gamma_1~\rightarrow ~\Gamma_4$ to be 99 meV, which is relatively close to the INS value of 123 meV and introduced the idea of an antiferromagnetic exchange interaction between the Pu$^{4+}$ ions\cite{colarieti2002PRB}. Recently, $Pegg$ $et$ $al.$\cite{pegg2018PCCP,pegg2018JPCC} suggested a longitudinal $3k$ AFM structure that retains the $Fm\overline{3}m$ symmetry and the calculated band gap (2.97 eV) associated with this magnetic structure agrees well with experimental optical band gap of 2.8(1) $eV$ at room temperature\cite{mccleskey2013JAP}, indicating the need to consider noncollinear and spin–orbit interactions in PuO$_2$. Furthermore, the first principles local density + $U$ (LDA+U) and generalized gradient + $U$ (GGA+U) approximations have also predicted an AFM ground state for PuO$_2$\cite{sun2008JCP,jomard2008PRB}.

\subsubsection{Americium Dioxide: AmO$_2$} 


AmO$_2$ crystallizes in the same $fcc$ fluorite structure as the other actinide dioxides, with a lattice parameter of $5.379$\AA \cite{nishi2008JNM}. Although AmO$_2$ is a simple tetravalent oxide isostructural to UO$_2$ and NpO$_2$, the magnetic ground state of AmO$_2$ is poorly understood due to the rarity of its parent actinide, along with the more serious limiting factor with experiments, the strong self-radiation-induced crystalline disorder that arises due to the continuous $\alpha-$particle decay of Am\cite{tokunaga2014PRB}. Despite these difficulties, more than 40 years ago, a phase transition was observed near $T_0$ = 8.5 K through magnetic susceptibility experiments demonstrating an effective moment of 1.31 $\mu_B$ and suggesting the ground state of Am$^{4+}$ to be a $\Gamma_7$ doublet carrying a dipole degree of freedom\cite{abraham1971PRB,kolbe1974JCP}. At the time, a comparison was made with the antiferromagnetic order seen in UO$_2$; however, M\"{o}ssbauer\cite{kalvius1969PLB} and neutron diffraction measurements failed to detect any AFM order\cite{boeuf1979JDPL}. More recently, the $\Gamma_7$ doublet has been shown to be converted to a $\Gamma_8$ quartet through spin-orbit couplings and Coulomb Interactions\cite{hotta2009PRB}. Since, the $\Gamma_8$ quartet can carry not only dipole but also quadrupolar and octupolar moments, it hints at the possibility of multipolar ordering in AmO$_2$\cite{hotta2009PRB,suzuki2013PRB}. 

The first and subsequent NMR measurements on AmO$_2$ done by $Tokunaga$ $et$ $al.$ confirmed the phase transition seen at $T_0 = 8.5$ K\cite{tokunaga2014PRB,tokunaga2010JPSJ}. The signal wipeout below $12$ K and a broadening of the spectrum with a randomly distributed hyperfine field developing at the lowest temperature of $1.5$ K\cite{tokunaga2014PRB} revealed short-range, spin-glass-like character for the magnetic phase transition seen at $T_0$. The NMR data was remarkably similar to that of a geometrically frustrated antiferromagnetic NiGa$_2$S$_4$\cite{matsuura2003PRB}. Over the course of the experiment, significant NMR intensity change occurred, indicating the electronic ground state is extremely sensitive to disorder induced by self-irradiation through the decay of Am\cite{tokunaga2014PRB}. Moreover, the NMR spectrum of AmO$_2$ is rather different from that seen in the antiferromagnetic ground state of UO$_2$ or the multipolar ground state of NpO$_2$\cite{tokunaga2011JPSJ}. The intense radioactivity and self-induced disorder present in any AmO$_2$ powder sample make experimentation particularly delicate and, thus, limit the available experimental techniques. Hence, detailed characterization on AmO$_2$ is severely lacking. Single crystal neutron scattering, NMR, Muon spin relaxation ($\mu$SR), or even a specific heat study could help reveal the exact nature of the phase transition observed in AmO$_2$.

\subsubsection{Curium Dioxide: CmO$_2$} 

Like PuO$_2$ and AmO$_2$, CmO$_2$ is lacking in experimental data, and the existing data sometimes contradict one another. CmO$_2$ crystallizes in the shared $fcc$ fluorite structure with a lattice constant of $5.359(2)$ \AA \cite{peterson1971JINC}.  In theory, the Cm$^{4+}$ ion with a 5$f^6$ configuration in CmO$_2$ should exhibit a singlet ground state with $J=0$ and therefore, be nonmagnetic. Yet, magnetic susceptibility measurements of CmO$_2$ show a Curie-Weiss behavior (instead of temperature independence) and a large effective moment of 3.36(6) $\mu_B$\cite{nave1983PRB,morss1989JLCM}. $Morss$ $et$ $al.$\cite{morss1989JLCM} determined that no long-range order was present in CmO$_2$ using neutron diffraction measurements, indicating the absence of any oxygen superlattice or superstructure and carefully determined the sample was stoichiometric CmO$_2$ without additional phases.

$Niikura~et~al.$\cite{niikura2011PRB} predicted the excitation energy between the ground and magnetic excited states become small due to the combined effect of Coulomb interactions, spin-orbit coupling, and crystal electric field potential and could be responsible for the magnetic behavior of CmO$_2$. The lattice parameters of CmO$_2$ deviate from the actinide contraction (a steady decrease in lattice parameters as the atomic number increases in the Actinide Series), suggesting a possible mixed-valence configuration for Cm atoms in CmO$_2$\cite{morss2006Springer,huang2020PRB}. Recently, $Huang$ $et$ $al.$\cite{huang2020PRB} theorized the Cm ions in CmO$_2$ consist of such a mixed-valence electronic structure between the 5$f^6$ and 5$f^7$ configurations. The Cm$^{3+}$ ion with $5f^{7}$ could contribute a moment up to 7.94$\mu_B$. In this mixed-valence scenario, the effective magnetic moment agrees well with the experimental value\cite{huang2020PRB}. However, these theoretical models for CmO$_2$ have not been confirmed by experiments. Making matters worse, CmO$_2$ does not appear to be an insulator with a large band gap like the other actinide oxides; rather, the band gap is predicted to be small, on the order of $\sim0.4$ eV \cite{prodan2007PRB,hou2017PRB}. High-temperature magnetic susceptibility and X-ray absorption spectroscopy experiments may provide further insight into the mixed-valence scenario suggested for CmO$_2$.

\subsubsection{Berkelium Dioxide: BkO$_2$} 

Following the discovery of americium and curium, berkelium (Bk) and californium were produced and discovered via nuclear bombardment in 1949-50. However, due to the difficulty in production, the most stable isotope of Berkelium, $^{247}$Bk is not commonly used for experimental work; rather the $^{249}$Bk isotope is the only isotope available in bulk form, and thus is used for experimental studies \cite{putkov2021RJPCA,white2019CEJ}.  BkO$_2$ also crystallizes in the same $fcc$ fluorite structure shared by the other actinide dioxides with a lattice parameter of $5.334(5)$ \AA \cite{baybarz1968JINC}. In BkO$_2$, berkelium exhibits the same $4+$ oxidation state as the other actinide dioxides, therefore exhibiting a half-filled $5f^7$ electronic configuration with a total angular momentum $L = 0$.

Early electron paramagnetic resonance (EPR)\cite{boatner1978RPP} and magnetic susceptibility\cite{karraker1975JCP} experiments determined the ground state to be the $\Gamma_6$ doublet, while the $\Gamma_8$ quartet followed by $\sim 6-10$ meV. These measurements used BkO$_2$ in a ThO$_2$ matrix. The magnetic susceptibility displayed two temperature-dependent paramagnetic regions with effective moments of $7.66~ \mu_B$ ($10$ to $95$ K) and $5.8~\mu_B$ ($95$ to $220$ K) while below $10$ K, the data suggests a possible AFM transition at $3$ K\cite{karraker1975JCP}. Later, $Nave~et~al.$\cite{nave1983PRB} reported an effective moment of $7.92~ \mu_B$ (a localized $5f^7$ configuration) as found by a Curie-Weiss fit to the magnetic susceptibility (50-300 K) with a large Weiss temperature~$\theta_W$ =  $250(50)$ K. Calculating the crystal field levels using the perturbative model proposed by $Magnani$ $et$ $al.$, finds agreement with the ground state levels. However, the energy splitting is significantly smaller ($\sim 1$ meV) than experimental estimates (6-10 meV)\cite{boatner1978RPP,karraker1975JCP,magnani2007JPCS}. Recently, $Putkov$ $et$ $al.$\cite{putkov2021RJPCA} calculated the electronic structure and XPS spectrum for the valence electrons of BkO$_2$ in the $0$ to $50$ eV range, finding excellent agreement with the experimental XPS spectrum\cite{veal1977PRB} and suggesting covalence between orbitals plays a major role in BkO$_2$. Neutron diffraction, NMR or $\mu$SR measurements are critical to resolve the nature of the phase below the $T$ = 3 K magnetic susceptibility phase transition.

\subsubsection{Californium Dioxide: CfO$_2$} 

Californium, like berkelium, is one of the last elements to be discovered in the actinide series. CfO$_2$ crystallizes in the same $fcc$ fluorite structure shared by the other actinide dioxides with a lattice parameter of 5.310(2) \AA \cite{baybarz1972JINC}. Californium can adopt a range of oxidation states ranging from $2+$ to $5+$\cite{kovacs2018IC} and in CfO$_2$, californium exhibits a $4+$ oxidation state, yielding a $5f^8$ electronic configuration and a $J=6$ multiplet. Magnetic susceptibility experiments determined that this multiplet is heavily split by crystal electric fields between $20$ and $100$ K and CfO$_2$ exhibited antiferromagnetic ordering below $T_N$ = 7 K\cite{moore1986JLCM}. Above $100$ K, CfO$_2$ displays Curie-Weiss behavior with an effective moment $\mu_{eff}$ = 9.1(2) $\mu_B$, in agreement with the theoretical moment of 9.3 $\mu_B$\cite{moore1986JLCM}. $Magnani$ $et$ $al.$\cite{magnani2007JPCS} proposed a perturbative model for the crystal electric fields in the actinide elements and determined the ground state of CfO$_2$ is the $\Gamma_3$ doublet with one of the $\Gamma_5$ triplets laying only $3.5$ meV higher. The calculated magnetic susceptibility of this perturbative model agrees exceptionally well with existing experimental magnetic susceptibility data\cite{moore1986JLCM}. 

\subsubsection{PaO$_2$} 
Protactinium (Pa) has no current uses outside of scientific research, and in combination with its activity, rarity, and toxicity, it is one of the least studied and understood actinide elements. While protactinium can form various compounds with oxidation states ranging from $2+$ to $5+$, it is most commonly present in the $5+$ oxidation state where it forms the pentoxide Pa$_2$O$_5$. Reduction of the pentoxide using hydrogen at 1550 $^\circ$C forms the dioxide PaO$_2$, where Protactinium takes a $4+$ oxidation state with a $5f^1$ electronic configuration\cite{sellers1954JACS}. PaO$_2$ crystallizes in the $fcc$ structure with a lattice constant of 5.505(1) \AA \cite{sellers1954JACS}. Recently, $Obodo$ $et$ $al.$ predicted PaO$_2$ to be a Mott-Hubbard insulator with a band gap of 3.48 eV\cite{obodo2013JPCM}; however, no experimental studies have been made for comparison.

\subsection{Magnetic properties of sesquioxides}\label{sec4b}

\subsubsection{Cf$_2$O$_3$} 

In the lanthanide sesquioxides, three crystal structures are known (the M$_2$O$_3$ structures): the body-centered cubic ($bcc$) Mn$_2$O$_3$ structure, the hexagonal La$_2$O$_3$ structure, and the monoclinic Sm$_2$O$_3$ structure. Cf$_2$O$_3$ is known to crystallize in the monoclinic Sm$_2$O$_3$ type (lattice parameters $a=14.124(20)$ \AA, $b=3.591(3)$ \AA, $c=8.809(13) $\AA)\cite{green1965UCB}, as well as the cubic Mn$_2$O$_3$ structure (lattice parameter $a=10.839(4) $\AA) \cite{morss1987JLCM}. Californium exhibits a $3+$ oxidation state in the sesquioxide Cf$_2$O$_3$, yielding a 5f$^9$ electronic configuration. Both cubic and monoclinic structures of Cf$_2$O$_3$ exhibit Curie-Weiss behavior above $100$ K with effective moments $\mu_{eff} = 9.8(2) \mu_B$ ($bcc$) and $\mu_{eff} = 10.2(2) \mu_B$(monoclinic)\cite{moore1986JLCM}. Moore et al.\cite{moore1986JLCM} reported both structures exhibit AFM ordering with $T_N$ = 19.0(1.5) K for cubic and $T_N$ = 8(2) K for monoclinic structure, while in between $20$ and $100$ K, the magnetic susceptibilities displayed pronounced crystal field effects. However, later Morss et al.\cite{morss1987JLCM} reported no magnetic ordering down to $1.5$ K based on their magnetic susceptibility measurements on $bcc$ Cf$_2$O$_3$. Morss et al.\cite{morss1987JLCM} noted that their study used $1.23$ mg sample for magnetic susceptibility measurements, which is significantly larger than the $31$ $\mu$g sample used by Moore et al.\cite{moore1986JLCM} and the magnetic ordering could have been impacted due to significantly more radioactive self-heating. The magnetism of Cf$_2$O$_3$ is not yet fully resolved and single crystal neutron diffraction, magnetic susceptibility, or $\mu$SR studies could resolve the magnetic ordering observed in Cf$_2$O$_3$. 

\subsubsection{Cm$_2$O$_3$} 

Cm$_2$O$_3$ exhibits all three known M$_2$O$_3$ structure types as well as two additional high-temperature polymorphic phases\cite{turcotte1973JINC}; above $1615$ $^\circ$C Cm$_2$O$_3$ forms the hexagonal La$_2$O$_3$ structure, between $800-1615$ $^\circ$C the monoclinic Sm$_2$O$_3$ structure forms, and below $800$ $^\circ$C, the $bcc$ Mn$_2$O$_3$ structure forms\cite{morss1983IC,konings2001JNM}. Lattice parameters for the hexagonal structure are $a$ = $3.80(2)$\AA, $c$ = $6.00(3)$ \AA\: at 1750 $^\circ$C, \cite{wallmann1964JINC,rimshaw1969curium}, for the monoclinic structure are $a$ = 14.282(8)\AA, $b$ = 3.652(3) \AA, $c$ = 8.900(5)\AA, $\beta$ = 100.31(5)$^\circ$\cite{haug1967JINC}, and for the $bcc$ structure $a$ = 10.97(1)\AA\cite{wallmann1964JINC}. In Cm$_2$O$_3$, curium exhibits the same $3+$ oxidation state as the other sesquioxides, therefore exhibiting a half-filled $5f^7$ configuration with total angular momentum $L = 0$. So, Cm$_2$O$_3$ could exhibit the $\Gamma_6$ doublet ground state like in BkO$_2$; however, this has yet to be verified. 


Magnetic susceptibility experiments determined monoclinic Cm$_2$O$_3$ exhibits Curie-Weiss behavior between $100-300$ K with $\mu_{eff}$ = 7.89(4)$\mu_B$ whereas the calculated moment for the free ion is 7.94 $\mu_B$\cite{morss1983IC}. Furthermore, Morss et al.\cite{morss1983IC} reported low-temperature antiferromagnetic ordering below 13(2) K\cite{morss1983IC}. Sample preparation using the very long-lived $^{248}$Cm (half-life = 3.39x10$^5$ year) isotope over the high $\alpha$ specific activities of the most common $^{242}$Cm (half-life = 163 days) isotope could enable further physical property characterization on this compound beyond the known current crystal structures\cite{morss1983IC}. Neutron diffraction, $\mu$SR, and specific heat studies are needed to further understand magnetic properties of Cm$_2$O$_3$. 


\subsubsection{Am$_2$O$_3$} 

Am$_2$O$_3$, like its dioxide AmO$_2$, is very poorly understood. Like the other actinide sesquioxides, Am$_2$O$_3$ is known to crystallize in the three M$_2$O$_3$ structures: the $bcc$ Mn$_2$O$_3$ structure, the hexagonal La$_2$O$_3$, and the monoclinic Sm$_2$O$_3$ structure. Lattice parameters for the hexagonal structure are $a=3.8123(3)$ \AA, $c=5.9845(5)$ \AA \cite{hurtgen1977INCL,Horlait2014JSSC,chikalla1968JINC}, for the monoclinic structure are $a=14.315(2)$ \AA, $b=3.6793(5)$ \AA, $c=8.927(1)$ \AA, $\beta=100.37(2)^\circ$ \cite{chikalla1968JINC,Horlait2014JSSC}, and for the $bcc$ structure $a=11.012(5)$ \AA \cite{hurtgen1977INCL,chikalla1968JINC}. The monoclinic Am$_2$O$_3$ is reported to have O/Am ratios between 1.54 and 1.51\cite{chikalla1968JINC} and has never been isolated alone\cite{noutack2019PRM}. 

XAFS measurements\cite{nishi2010JNM} show excellent agreement with existing X-ray diffraction data and theoretical predictions using the Full-Potential Linearized Augmented Plane Wave (FP-LAPW) method in DFT, indicating the electronic structure and formation enthalpies of hexagonal Am$_2$O$_3$ can be modeled successfully using DFT\cite{noutack2019PRM,konings2014JPCRD,hurtgen1977INCL,nishi2010JNM}. Recently, modeling using the GGA+U as well as hybrid functionals has predicted hexagonal Am$_2$O$_3$ to be a Mott insulator with a band gap around $~2.85$ eV (U = $5$ eV) and determined it is the most stable form at low temperatures\cite{suzuki2012JPCS,noutack2019PRM}. Unlike AmO$_2$, americium in Am$_2$O$_3$ exhibits a $3+$ oxidation state in the sesquioxide, yielding a 5f$^6$ electronic configuration with $J$ = 0. Thus, Am$_2$O$_3$ may have a non-magnetic ground state yet to be experimentally verified.

\subsubsection{Pu$_2$O$_3$} 

$\beta$-Pu$_2$O$_3$ crystallizes in the hexagonal La$_2$O$_3$ structure shared with the other actinide sesquioxides as well as Nd$_2$O$_3$\cite{rai2020PRB,sala2018PRM}, with lattice parameters $a$ = 3.841(6) \AA and $c$ = 5.958(5)\AA\cite{gardner1965JINC}. In Pu$_2$O$_3$, plutonium exhibits a ${3+}$ oxidation state, yielding a $5f^5$ electronic configuration with a Kramer's doublet ground state. A heat capacity anomaly associated with an antiferromagnetic transition $T_N$ = 19 K\cite{mccart1981JCP} was measured by Flotow et al.\cite{flotow1981JCP}. Mccart et al.\cite{mccart1981JCP} reported two magnetic transitions ($T_N$ = 19 K and $T$ = 4 K) in $\beta$-Pu$_2$O$_3$ from neutron diffraction and magnetic susceptibility measurements. Later Wulff et al.\cite{wulff1988JCP} solved the magnetic structures below both magnetic transitions by performing neutron powder diffraction. The magnetic structure below $T_N$ = 19 K was described by a magnetic cell that is doubled relative to the chemical unit cell in all three crystallographic directions and below 4 K the magnetic cell is identical to the chemical unit cell. The magnetic moment was found to be 0.60(2) $\mu_B$ per Pu ion pointing along the unique $c$ axis at all temperatures below $T_N$\cite{wulff1988JCP} and is consistent with a Kramer's doublet ground state.  

In addition to the stoichiometric $\beta$-Pu$_2$O$_3$, two forms of sub-stoichiometric $\alpha$-Pu$_2$O$_3$ sesquioxides are known to exist due to loss of oxygen from the dioxide. The first form, $\alpha$-Pu$_2$O$_3$, has an O/Pu ratio of $1.515$ with a $bcc$ structure with lattice parameter $a$ = 11.04(2)\AA. The second form, $\alpha'$-Pu$_2$O$_3$, has an O/Pu ratio of $1.61$ with a $fcc$ structure with lattice parameter $a$ = 5.409(1)\AA \cite{gardner1965JINC}. A theoretical DFT + $U$ study predicts an AFM semiconducting ground state for $\alpha$-Pu$_2$O$_3$ with an electronic band gap of $1.40$ eV\cite{lu2014PLA}, which has yet to be experimentally measured.

\subsection{Magnetic properties of other actinide oxides}\label{sec4c}

\subsubsection{U$_2$O$_5$} 

U$_2$O$_5$ lies on the transition point between the fluorite structure and the layered structures found in other uranium oxides and is one of the least understood of the various uranium oxides, owing to its instability relative to the other oxides\cite{molinari2017IC}. U$_2$O$_5$ is known to crystallize in at least 4 structures, $\alpha$-, $\beta$-, $\gamma$-, and $\delta$-, indicating complicated structures of U$_2$O$_5$\cite{hoekstra1970JINC,kovba1979R}. The crystal symmetry of $\alpha$-U$_2$O$_5$ is unknown (Unk.), while $\beta$-U$_2$O$_5$ is hexagonal with lattice parameters $a=b=3.813$ \AA, $c=13.180$ \AA, and the $\gamma$-U$_2$O$_5$ is monoclinic with $a=5.410$ \AA, $b=5.481$ \AA, $c=5.410$ \AA, $\beta=90.49^\circ$ \cite{hoekstra1970JINC}. $\delta$-U$_2$O$_5$ crystallizes in the orthorhombic structure ($Pnma$ space group) with $a=6.849$ \AA, $b=8.274$ \AA, and $c=31.706$ \AA \cite{kovba1979R} and contains a mixture of distorted O coordination geometries (6- and 7-fold). 

Recently, $\delta$-U$_2$O$_5$ was successfully modeled using the DFT+U method suggesting only U$^{5+}$ to be present\cite{brincat2015DT}. All U$_2$O$_5$ structures, $\alpha$-, $\beta$-, $\gamma$-, and $\delta$-, are predicted to be insulators with band gaps of $2.01$, $2.23$, $2.19$, and $1.60$ eV, respectively\cite{brincat2015DT}. DFT has also shown $\delta$-U$_2$O$_5$ to have the highest formation energy of any known uranium oxide \cite{andersson2013IC}. Using the GGA+U method in DFT predicted that Np$_2$O$_5$ structured U$_2$O$_5$ is just slightly more stable than $\delta$-U$_2$O$_5$ under ambient conditions and no thermodynamic studies have been reported for it, suggesting a new avenue of work in low-temperature uranium oxides\cite{molinari2017IC}. 

\subsubsection{Np$_2$O$_5$} 

NpO$_2$ and Np$_2$O$_5$ are the only two stable oxides of neptunium. NpO$_2$ has been extensively studied from  both experimental and theoretical studies compared to Np$_2$O$_5$ and studies on Np$_2$O$_5$ are severely lacking\cite{forbes2007JACS,yun2011PRB,zhang2018JNM}. Forbes et al.\cite{forbes2007JACS} reported the structure of Np$_2$O$_5$ to be monoclinic ($P2/c$ Space Group) with lattice parameters $a$ = 8.168(2)\AA, $b$ = 6.584(1)\AA, $c$ = 9.313(1) \AA\:and $\beta$ = 116.09(1)$^{\circ}$ and to contain three symmetrically distinct neptunyl(V) ions.

In Np$_2$O$_5$, neptunium exhibits a $5+$ oxidation state, yielding a $5f^2$ electronic configuration. Magnetic susceptibility data on a polycrystalline sample revealed a cusp at $22(3)$ K, suggesting long-range antiferromagnetic ordering. An effective moment of 2.2(1) $\mu_B$ and a Weiss constant of $-43(5)$ K were extracted from the Curie-Weiss fit\cite{forbes2007JACS}. The effective moment is significantly smaller than the free ion value of $3.58$ $\mu_B$ and may result from crystal-field effects on the $f$-spin states. Np$_2$O$_5$ could exhibit complex magnetic structures due to the three distinct Np crystallographic sites and the Np-O-Np superexchange pathways, which are impacted by bond lengths and angles. Yun et al.\cite{yun2011PRB} predicted Np$_2$O$_5$ to have insulting behavior and complicated noncollinear magnetic order with a ferromagnetic (FM) exchange coupling along the $c$ axis and a weaker antiferromagnetic coupling in the $a-b$ plane. Experiments like neutron diffraction and magnetic susceptibility on Np$_2$O$_5$ single crystals are essential to understand the exact magnetic ground state and examine the impact of multiple Np sites and the Np-O-Np superexchange pathway in generating competing exchange interactions leading to long-range magnetic order.

\subsubsection{Pa$_2$O$_5$} 

Despite Pa$_2$O$_5$ being the most stable protactinium oxide, the ground state and crystal structure are still poorly understood. The only experimental study on Pa$_2$O$_5$ was undertaken by $Sellers$ $et$ $al.$ in 1954, and reported not only an $fcc$ structure with a lattice constant of 5.455(7) \AA~but a second, layered orthorhombic structure, isostructural with Nb$_2$O$_5$ and Ta$_2$O$_5$ pentoxides\cite{sellers1954JACS}.~In this orthorhombic structure, the unit cell was found to be $6.92(2)$\AA~by $4.02(1)$\AA~by $4.18(2)$\AA\cite{sellers1954JACS}. In its cubic form, Pa$_2$O$_5$ follows the fluorite cubic structure with the additional oxygen atoms randomly occupying available holes \cite{sellers1954JACS}. Recently, $Liu$ $et$ $al.$\cite{isbill2022PRM} has shown that doping excess oxygen atoms into the octahedral position of Protactinium is the most stable doping position for the fluorite structure and even predicted Pa$_2$O$_5$ to crystallize in the $\zeta$-Nd$_2$O$_5$ structure, resulting in a charge-transfer insulator with a band gap of $2.67$ eV. This $\zeta$-Pa$_2$O$_5$ phase has not been experimentally observed; however due to the aforementioned lack of protactinium oxide investigations it remains a possibility. Recent work on the excitation spectrum of protactinium has shown significant underestimations of the density of states, which needs to be considered in any future theoretical work on the protactinium oxides\cite{naubereit2018PRA}. The synthesis of single crystals of Pa$_2$O$_5$ and performing single crystal X-ray diffraction (XRD) are key to resolving the existing ambiguity in the exact crystal structure. Furthermore, the availability of high-quality crystals will provide an opportunity to probe the thermodynamic properties of Pa$_2$O$_5$.

\begin{figure}[t!] 
\includegraphics[clip,width=\columnwidth]{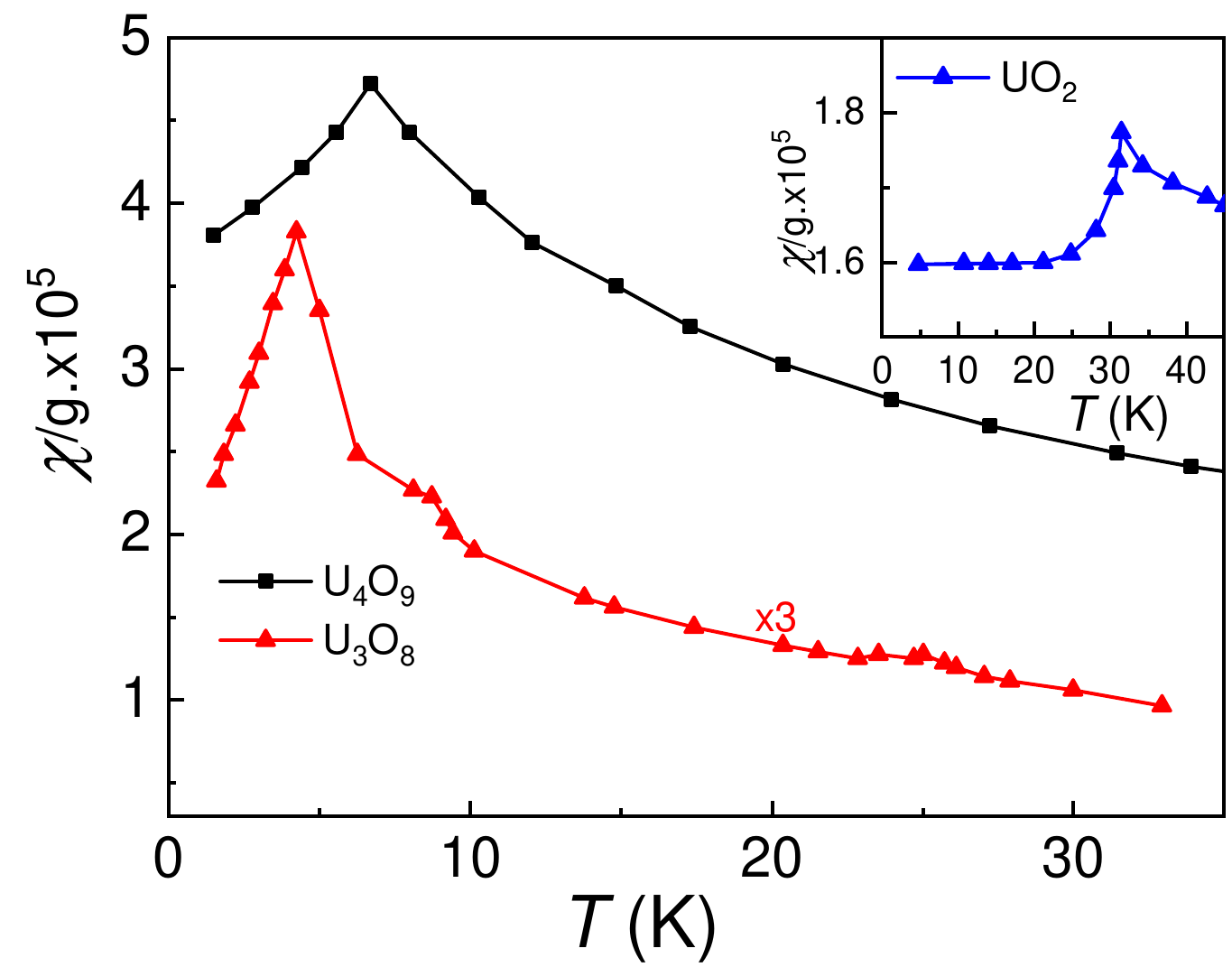}
\caption{\label{U_sus} Magnetic susceptibility per g of  UO$_2$ (in inset)\cite{leask1963JCS},  U$_3$O$_8$\cite{leask1963JCS} and U$_4$O$_9$\cite{leask1963JCS} as a function of temperature showing a cusp anomaly related to a magnetic phase transition. Note: Magnetic susceptibility of U$_3$O$_8$ is multiplied by 3 times the original susceptibility values for the better visibility of the phase transitions within the frame.}
\end{figure}

\subsubsection{U$_3$O$_8$}

Triuranium octoxide, U$_3$O$_8$, can be formed through oxidation of UO$_2$ and is one of the most stable forms of uranium under environmental conditions. U$_3$O$_8$ is reported to crystallize in two common polymorphs ($\alpha$-U$_3$O$_8$ and $\beta$-U$_3$O$_8$). At room temperature, $\alpha$-U$_3$O$_8$ crystallizes in an orthorhombic $Amm2$ structure while upon heating $\alpha$-U$_3$O$_8$ to 1350 $^{\circ}$C, an expansion along the $c$-direction along with an accompanying contraction along the $b$-direction induces an isomorphic transition in structure to the $P62m$ phase at $573$ K\cite{loopstra1970JAC,ackermann1977JINC,siegel1955AC}. The low temperature $Amm2$ $\alpha$-U$_3$O$_8$ phase has two distinct crystallographic sites, one U(VI) site and one double degenerate U(V) site (2U(V)), while the high temperature $P62m$ $\beta$-U$_3$O$_8$ phase has only a single crystallographic site\cite{yun2011PRB,ranasinghe2020CMS,brincat2015DT,miskowiec2021PRB}. $\alpha$-U$_3$O$_8$ is semiconducting with a band gap of $\approx 1.80$ eV\cite{ranasinghe2020CMS}.

Early magnetic susceptibility experiments displayed three distinct peaks, a single large peak at $4.2$ K, and two other small peaks at $8$ and $25.3$ K, as shown in Fig.\:\ref{U_sus} \cite{leask1963JCS}. Heat capacity measurements also displayed a peak at $25.3$ K\cite{westrum1959JACS}.  Recent INS experiments confirmed AFM below $22$ K (onset between $25$ and $22$ K) down to at least $1.7$ K, with peaks corresponding to $[0.5$ $1$ $1]$ and [$0.5$ $2$ $2$] in an orthorhombic structure\cite{miskowiec2021PRB}. No magnetic intensity was observed at [$0.5$ $1$ $1$], ruling out AFM order along the $a$ axis\cite{wen2012JPCM,miskowiec2021PRB}. A quasielastic scattering term appears above $100$ K, increasing in intensity and width up to $600$ K, confirming $\alpha$-U$_3$O$_8$ is a mixed-valence state system, similar to CmO$_2$\cite{miskowiec2021PRB}. $Isbill$ $et$ $al.$\cite{isbill2022PRM} recently used DFT to show that antiferromagnetic ordering along the $[0.5$ $1$ $1]$ is indeed lower in energy than along any Miller indices. $Isbill$ $et$ $al.$\cite{isbill2022PRM} determined that moments which point along the in-plane directions will gradually relax until they point along a direction between the two distinct crystallographic uranium sites in rather complex noncollinear structures.


\subsubsection{U$_4$O$_9$}

U$_4$O$_9$ is formed in the process of oxidizing UO$_2$ to U$_3$O$_8$ and crystallizes in three structures ($\alpha$-, $\beta$-, and $\gamma$-) which are based on the fluorite crystal structure of UO$_2$.  $\alpha$- and $\beta$-U$_4$O$_9$ form similar super-structures to that of UO$_2$, but with a unit cell 4 times larger. In cubic $\beta$-U$_4$O$_9$, additional oxygen atoms are ``accommodated" in cuboctahedral clusters (space group $I43d$) whose centers are unoccupied\cite{bevan1986JSSC,popa2004ACSFC,cooper2004ACSFC}. In $\alpha$-U$_4$O$_9$ the same cuboctahedral clusters form, however, the positions are slightly displaced due to a shift in the position of the central atom (trigonal distortion) and are extremely similar to that of U$_3$O$_8$\cite{desgranges2011IC}. These clusters form layered sheets which transform driven by shear transformations from $\alpha$-U$_4$O$_9$ into U$_3$O$_8$ sheets\cite{desgranges2011IC,desgranges2016IC} and this transformation has been shown to be accompanied by a complex change in U oxidation states\cite{kvashnina2013PRL}.  Raman spectroscopy has recently determined a band at $630$ cm$^{-1}$ in U$_4$O$_9$ determined to be characteristic of clusters composed of the excess ``accommodated" interstitial oxygen atoms\cite{desgranges2012JRS}. Neutron diffraction measurements have shown $\beta-$U$_4$O$_9$ to be crystallographically ordered with no Uranium-Oxygen bonds shorter than $2.2$ \AA while $\alpha-$U$_4$O$_9$ was shown to have some Uranium-Oxygen bonds on the order of $1.8$ \AA \cite{garrido2006IC,desgranges2011IC}. 

Recent high energy resolution limited fluorescence detection X-ray absorption near edge spectroscopy (HERFD-XANES) measurements\cite{leinders2017IC} have shown uranium adopts a mixed valence configuration in U$_4$O$_9$, with U$^{4+}$ and U$^{5+}$ oxidation states\cite{desgranges2016IC}. The high temperature ($T$ = $77$ to $500$ K) magnetic susceptibility experiments determined paramagnetic behavior in U$_4$O$_9$\cite{gotoo1965JPCS}. The low-temperature ($T$ = $1.5$ to $44$ K) magnetic susceptibility measurements reported a maximum at $6.4$ K\cite{leask1963JCS}. However, the low-temperature phase transition at $6.4$ K was not observed in specific heat measurements\cite{osborne1957JACS,flotow1968JCP}. New magnetic susceptibility and specific heat measurements with improved experimental techniques and improved synthesis routes could resolve ambiguity around the low-temperature phase transition observed at $6.4$ K.

\subsubsection{U$_3$O$_7$}
U$_3$O$_7$ is another oxide formed during the oxidation of UO$_2$ to U$_3$O$_8$. Historically, U$_3$O$_7$ was thought to crystallize in two structures, $\alpha$- and $\beta$-U$_3$O$_7$. However, early experiments failed to differentiate $\alpha$-U$_3$O$_7$ from cubic $\beta$-U$_4$O$_9$ in the sample\cite{westrum1962JPCS,hoekstra1961JINC,rousseau2006JNM,leinders2016IC}. Neutron diffraction studies and DFT have confirmed U$_3$O$_7$ crystallizes in a fluorite-type structure consisting of cuboctahedral oxygen clusters very similar to U$_4$O$_9$\cite{cooper2004ACSFC,desgranges2011IC}; however, the clusters are tilted and skewed from those seen in U$_4$O$_9$\cite{desgranges2009IC,andersson2013IC,brincat2015JNM,leinders2016IC}. The fluorite-like structure can be characterized by a unit cell containing 15 fluorite-like subshells with lattice parameters $a=b=5.3900(2)$ \AA, and $c=5.5490(2)$ \AA ($c/a=1.031$)\cite{leinders2016IC}. Recent total reflection X-ray fluorescence-based XANES (TXRF-XANES) has shown U$_3$O$_7$ is a mixed-valence compound, consisting of uranium in both U$^{4+}$ and U$^{5+}$ states\cite{sanyal2017AC}. First principle calculations using the Vienna ab initio simulation (VASP with the projector-augmented wave (PAW) method) and DFT combined with $Perdew, Burke$, and $Ernzerhof$ functionals (PBE+U) to account for exchange-correlation interactions, including spin-orbit coupling, predicts U$_3$O$_7$ to be a charge-transfer insulator with a band gap of $0.32$ eV\cite{leinders2021IC}. Additionally, in such a structure with cuboctahedral clusters, U$^{4+}$ and U$^{5+}$ are predicted to carry noncollinear magnetic moments of $1.6$ and $0.8\mu_B$ respectively, resulting in the fluorite-based structure having canted FM order in characteristic layers\cite{leinders2021IC}. However, this FM order has yet to be experimentally observed.

\subsubsection{UO$_3$}
Heating uranyl nitrate to $400$ $^\circ$C forms Uranium trioxide, UO$_3$\cite{sheft1950JACS}, in which uranium takes a $6+$ oxidation state. The uranium atoms within UO$_3$ can be coordinated with either $6$, $7$, or $8$ oxygen atoms leading to at least $7$ known structures, $\alpha$-, $\beta$-, $\gamma$-, $\delta$-, $\varepsilon-$, $\zeta$-, and $\eta$- as well as an amorphous UO$_3$\cite{cordfunke1965TFS,loopstra1966RTCPB,debets1966AC,greaves1972ACB,loopstra1977JSSC,hoekstra1970JINC,siegel1966AC,weller1988Poly,cordfunke1988TA,sweet2013JRNC}. $\alpha$-UO$_3$ crystallizes in an orthorhombic structure ($C2mm$\cite{loopstra1966RTCPB} or $C222$\cite{greaves1972ACB} space group), while $\beta$-UO$_3$ crystallizes in a monoclinic structure\cite{debets1966AC} with lattice parameters $a=10.34$\AA, $b=14.33$\AA, $c=3.910$\AA, and $\beta=99.03^\circ$\cite{enriquez2020ACSAMI,debets1966AC}. $\gamma$-UO$_3$ crystallizes in an orthorhombic structure ($Fddd$ space group) at room temperature and lower the symmetry to a tetragonal ($I4_1$ space group) structure at high temperatures\cite{loopstra1977JSSC}. $\delta$-UO$_3$ crystallizes in the ReO$_3$ structure ($Pm3m$ space group)\cite{weller1988Poly}. $\varepsilon$-UO$_3$ crystallizes in a triclinic ($P-1$) structure with lattice parameters $a=4.01$ \AA, $b=3.85$ \AA, $c=4.18$\AA, $\alpha = 98.26 ^\circ$, $\beta = 90.41 ^\circ$, and $\gamma = 120.46 ^\circ$, best described by a $2$x$1$x$2$ supercell of  $P-1$ with lattice parameters $a = 8.03$ \AA, $b = 3.86$ \AA, $c = 8.37$ \AA, and $\beta = 90.41 ^\circ$\cite{spano2022JNM}. Heat capacity and susceptibility measurements on high purity $\beta$- and $\gamma$-UO$_3$ suggest weak paramagnetism; however no anomalies were found between $1.3-5$ K\cite{cordfunke1988TA}. A recent DFT+U study has predicted three new structures for UO$_3$ under $13$, $62$, and $220$ GPa of pressure, predicting a semiconducting phase that transforms under further pressure to semi-metal and metallic phases\cite{ma2021PRB}. DFT+U has predicted a band gap around $1.6-2.6$ eV for $\alpha$-UO$_3$, while $\beta$-, $\gamma$-, and $\delta$-UO$_3$ all have predicted band gaps of $\sim2-2.4$, $\sim2.4-2.8$, and $\sim2.1-2.2$ eV respectively\cite{ma2021PRB,ao2021JNM,brincat2014IC,casillas2017PRM}. Epitaxial film growths of $\alpha$-UO$_3$ allowed an indirect measurement of the band gap of $2.26$ eV\cite{enriquez2020ACSAMI}.

\section{Summary and Outlook}\label{sec5}

The present article focuses on synthesis methods and bulk magnetic properties of the known actinide oxides, binaries in particular. We have also included a discussion of their crystal structures and appropriate methods for crystal growth. Since these actinide oxides are insulators and crystallize in relatively simple crystal structures, it might be expected that these compounds and their magnetic behavior would be well understood, both theoretically and experimentally. Contrary to this expectation, however, the magnetic ground states of actinide oxides, other than maybe ThO$_2$ and UO$_2$, are still unresolved due to the lack of high-quality samples, especially single crystals, the dual nature of 5$f$ electrons, and the electronic correlations and strong spin-orbit coupling associated with these actinide materials. 

An improved understanding of the chemical and physical properties of these actinide materials at low temperature may prompt new areas of actinide research, opening critical avenues for the design and control of novel phenomena related to 5$f$ electrons, including electronic correlations and topology. Moreover, this improved foundational understanding will impact advanced  modeling and materials design for high-temperature environments for nuclear industry applications\cite{zhang2012Science,pegg2018PCCP}. As detailed in our discussion on crystal growth, a lack of effort and progress in the growth of actinide oxide single crystals hinders the study of their fundamental structural and electronic properties, which are necessary for enabling the above opportunities. For example, U$_3$O$_8$ forms through further oxidation of UO$_2$, and studying the physical and chemical properties of high-quality crystals of U$_3$O$_8$ will allow us to gain new knowledge concerning its magnetic ground state, permitting us to develop models for interpreting and explaining uranium chemistry at a fundamental level, and ultimately, enable us to apply such an improved understanding to uranium science and research in the nuclear fuel industry. Therefore, the pursuit of high-quality samples is imperative for unraveling the intricate interplay of factors, including crystal field and strong spin-orbit coupling, which underlie the captivating magnetic behaviors observed in actinide oxides such as UO$_2$, CfO$_2$, CmO$_2$, Am$_2$O$_3$, and U$_3$O$_8$. Combining efforts of high-quality samples, detailed crystal structures, and magnetic characterization techniques holds the key to unraveling the complexities of magnetic ordering in these materials.

In general, the magnetic properties of the actinide dioxides appear to lack notable trends, however both theoretical models and existing experimental data hint at the possibility of similar ground states common to the various dioxides. For instance, experimental evidence from magnetic susceptibility and X-ray resonant scattering for NpO$_2$ suggests a magnetic phase transition at $T_0$ = $25$ K,  yet magnetic order has not been observed in neutron diffraction studies. Magnetic susceptibility measurements suggest a phase transition at $T_0$ = $8.5$ K in AmO$_2$, and confirmed through NMR measurements, which suggest short range, glass-like character. However, the possibility of multipolar ordering has not yet been ruled out for AmO$_2$, as well as for NpO$_2$. UO$_2$ displays a first order antiferromagnetic transition at $T_N$ = $30.8$ K with a $3-k$ noncollinear magnetic structure below $T_N$. Ultimately, UO$_2$, NpO$_2$, and AmO$_2$ might all be reasonably well described by various complicated magnetic structures or multipolar ordering. BkO$_2$ was suggested to exhibit an AFM transition at $T$ = 3 K, while an AFM transition at $T_N$ = $7$ K in CfO$_2$ was confirmed. In theory, CmO$_2$ should be nonmagnetic; however, it also displays a temperature-dependent magnetic susceptibility, hinting at the possibility of a multivalent configuration present in this material.  Despite the expected ground state for CmO$_2$ that should be nonmagnetic, it follows the trend set by the heavier BkO$_2$ and CfO$_2$, which both display Curie-Weiss behavior above $100$ K. However, whereas below $100$ K the magnetic susceptibilities of BkO$_2$ and CfO$_2$ become dominated by crystal field effects and subsequent AFM ordering, in CmO$_2$, the possibility of a multivalent configuration may complicate the crystal field effects and any subsequent order. 

The actinide sesquioxides and higher oxides have been largely understudied and the available data to date is too inconclusive to grasp at potential connections. Cf$_2$O$_3$ follows the trend set by its dioxide, exhibiting Curie-Weiss behavior above $100$ K with a large effective moment, and AFM ordering seen at $T_N$ = $19(1.5)$ K in the $bcc$ structure. That was later revisited with a larger sample suggesting no AFM order, together with $T_N$ = $8(2)$ K in the monoclinic structure. Cm$_2$O$_3$ again, follows the same trend, with an AFM transition at $T_N$ = $13(2)$ K. Am$_2$O$_3$ may be nonmagnetic; however no experimental measurements have been completed. Pu$_2$O$_3$ exhibits an AFM transition at $T_N$ = $19$ K, where the magnetic unit cell is double the chemical unit cell, and a second transition at $T$ = $4$ K, where the magnetic unit cell is equal to the chemical unit cell. Magnetic susceptibilities have never been measured for U$_2$O$_5$ and Pa$_2$O$_5$; however measurements for Np$_2$O$_5$ reveal an AFM transition at $22(3)$ K that might be explained by a complicated noncollinear magnetic structure which has yet to be verified experimentally. The higher uranium oxides, U$_3$O$_8$ and U$_4$O$_9$ display paramagnetic Curie-Weiss behavior at high temperatures and undetermined phase transitions at $4, 8,$ and $25.3$ K in U$_3$O$_8$ and $6.4$ K in U$_4$O$_9$, respectively. UO$_3$ displays weak paramagnetic behavior, while U$_3$O$_7$ has never undergone experimental magnetic characterization, although theory suggests noncollinear magnetic moments resulting in an unverified canted FM order.

At present, despite large previous experimental and theoretical efforts, a vast bounty of scientific knowledge regarding the actinide elements remains untapped, their applications and our understanding constrained by a dearth of needed synthetic and experimental work. It is our hope that this article will serve to alleviate at least some of the factors currently impeding such work, by promulgating further the requisite methods for single crystal syntheses and more broadly advertising the value of those syntheses. High-quality single crystals are critical for the determination of accurate crystal structures and investigations of complex magnetism and relevant electronic and thermodynamic properties, both in the chronically underexplored actinide oxides, as well as in actinide species more broadly. The understanding thus gained will play an indispensable role in unlocking new areas of actinide research (both applied and fundamental), and revealing the full breadth of possibilities the actinides present.

\section{Acknowledgments}

We thank Garret Gotthelf and Joseph Paddison for useful discussions and comments. B. K. R., A. B., and R. G. acknowledge support from the Laboratory Directed Research and Development (LDRD) program within the Savannah River National Laboratory (SRNL). This work was produced by Battelle Savannah River Alliance, LLC under Contract No. 89303321CEM000080 with the U.S. Department of Energy. Publisher acknowledges the U.S. Government license to provide public access under the DOE Public Access Plan (http://energy.gov/downloads/doe-public-access-plan). K.G. acknowledges support from the Division of Materials Science and Engineering, Basic Energy Sciences, Office of Science of the U.S. Department of Energy  (U.S. DOE). GM and HzL acknowledge support by the Center for Hierarchical Waste Form Materials (CHWM), an Energy Frontier Research Center (EFRC),  supported by the U.S. Department of Energy, Office of Basic Energy Sciences, Division of Materials Sciences and Engineering under Award DE-SC0016574. 



\bibliographystyle{unsrt}
\bibliography{Ref}

\end{document}